\documentclass[a4paper,11pt]{article}
\pdfoutput=1 
\usepackage{jheppub} 
\usepackage{subfigure}
\usepackage{caption}
\usepackage[T1]{fontenc}

\title{\boldmath Construction of two large-size four-plane micromegas detectors 
}

\author[a]{M.~Bianco, H.~Danielsson, J.~Degrange, R.~De Oliveira, E.~Farina, F.~Kuger, P.~Iengo$^{*}$, F.~Perez Gomez, G.~Sekhniaidze, F.~Sforza, O.~Sidiropoulou, M.~Vergain, J.~Wotschack}
\author[b]{A.~D\"udder, T.H.~Lin, M.~Schott$^{*}$, C.Valderanis}

\affiliation[a]{European Center for Nuclear Research, CERN, Geneva, Switzerland}
\affiliation[b]{Johannes Gutenberg-University, Mainz, Germany}

\emailAdd{matthias.schott@uni-mainz.de}
\emailAdd{paolo.iengo@cern.ch}

\abstract{We report on the construction and initial performance studies of two micromegas detector quadruplets with an area of 0.5\,m$^2$. They serve as prototypes for the planned upgrade project of the ATLAS muon system. Their design is based on the resistive-strip technology and thus renders the detectors spark tolerant. Each quadruplet comprises four detection layers with 1024 readout strips and a strip pitch of 415\,$\mu$m. In two out of the four layers the strips are inclined by $\pm$1.5$^{\circ}$ to allow for the measurement of a second coordinate. We present the detector concept and report on the experience gained during the detector construction. In addition an evaluation of the detector performance with cosmic rays and test-beam data is given. 
}

\begin{document} 
\maketitle
\flushbottom

\section{Introduction}

The upgrade of the Large Hadron Collider (LHC) to an instantaneous luminosity of 5$\times$10$^{34}$ cm$^{-2}$s$^{-1}$ in the years after 2020 will lead to an  increase of the particle rates in some of the LHC detectors in excess of what they have been designed for. In the ATLAS muon system \cite{ATLAS:MuonSpec} the first  station of the forward region is most affected. The expected occupancies in the currently installed Monitored Drift Tubes and Cathode Strip Chambers will be too high to allow for a reliable and precise reconstruction of muon tracks. It was therefore decided to replace these detectors and their support structures (Small Wheels) by the so-called New Small Wheels (NSWs) in 2018/19 \cite{Kawamoto:1552862}. 

The NSWs will be equipped with detectors based on two different technologies: sTGCs (multiwire proportional chambers) and micromegas detectors~\cite{Kawamoto:1552862}. Each detector type delivers precision track coordinates and trigger information, constituting a largely redundant trigger and tracking system. The NSWs will include eight layers of micromegas detectors, arranged in two quadruplets. In total, the NSWs comprises 128 quadruplets of 2-3m$^2$ size for a total active surface of more than 1200\,m$^2$. It represents the first system based on Micro Pattern Gaseous Detectors with such a large size.

\begin{figure}[tb]
	\centering
	\includegraphics[width=0.8\textwidth]{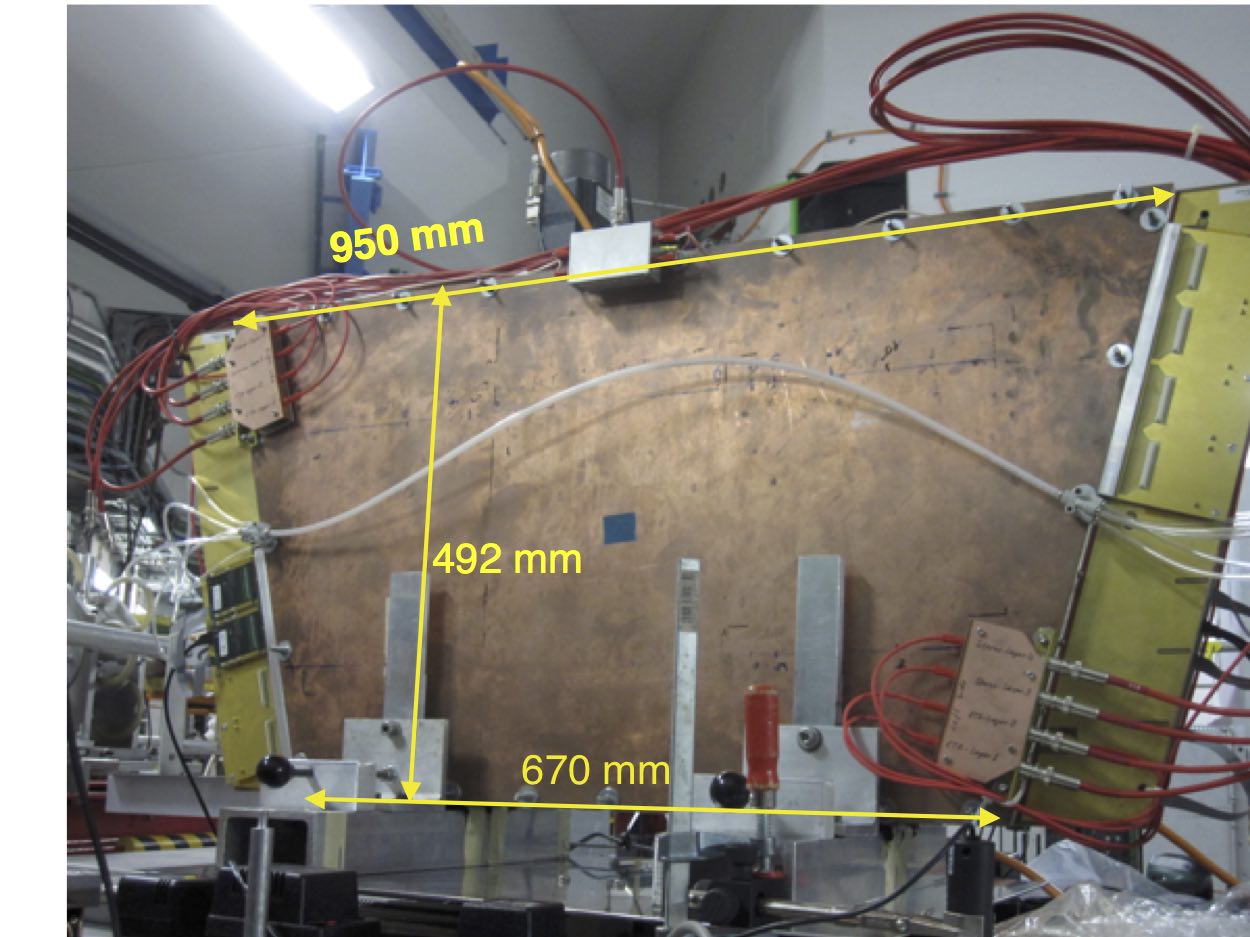}
	\caption{Photo of a MMSW detector being equipped with electronics. Along the sides the mezzanine cards are visible, however, not yet fully equipped with readout electronics, only the mezzanine card in the left lower corner carries two (out of the four) APV25 hybrid cards.}
	\label{Fig:FinalDetector}
\end{figure}

This article describes the construction of and the experience with two 0.5\,m$^2$ micromegas quadruplets (Fig.~\ref{Fig:FinalDetector}), called in the following MMSW\footnote{MicroMegas Small Wheel}. The construction of these detectors served the purpose of evaluating detector design, construction and performance issues relevant for the construction of the micromegas detectors to be installed on the NSWs. 

The paper is structured as follows. In Sec. \ref{sec:Layout} we discuss the design of the detector as well as details of the PCB boards. We give a step-by-step overview of the construction process, including also a discussion of the necessary tooling in Sec.~\ref{sec:Construction}. The results of first performance tests, based on cosmic rays, X-ray and test-beam measurements are discussed in Sec. \ref{sec:Performance}. The paper is summarized in Sec. \ref{sec:conclusion}.

\section{\label{sec:Layout}Detector design}

The MMSW detectors are meant to follow as much as possible the layout of the detectors on the NSWs, however, not full size. The dimensions of the MMSW were chosen such that the quadruplet can be installed on the existing Small Wheels behind the currently installed Cathode Strip Chambers. The latter are inclined by about 11$^\circ$ towards the interaction point, leaving, in the large sectors, between r=1.5 and 2~m just enough space to house a MMSW quadruplet. 

The detectors were designed to reach a spatial resolution of 100~$\mu$m in the precision radial ($\eta$) direction and 2--3~mm in the non-precision $\phi$ direction, following the requirements for the detectors to be installed in the NSWs. Tracks coming from the interaction point will traverse the MMSW detector, when installed on the present SW, under angles $\theta$ between 12$^\circ$ and 16$^\circ$. Any uncertainty in the z position on the detector plane translates with $\sin$($\theta$) into an uncertainty in the precision coordinate. The planarity of the detectors is, therefore, one of the major issues in the detector construction\footnote{This requirement becomes even more important for the NSW detectors with track angles up to 32$^\circ$.}. In the MMSW construction this issue has been addressed with the goal to achieve for each detection layer a planarity of better than 50~$\mu$m (rms) over the detector surface.

\subsection{Mechanics\label{sec:mechanics}}

\begin{figure}[htb]
\begin{minipage}[hbt]{0.49\textwidth}
	\centering
	\includegraphics[width=0.98\textwidth]{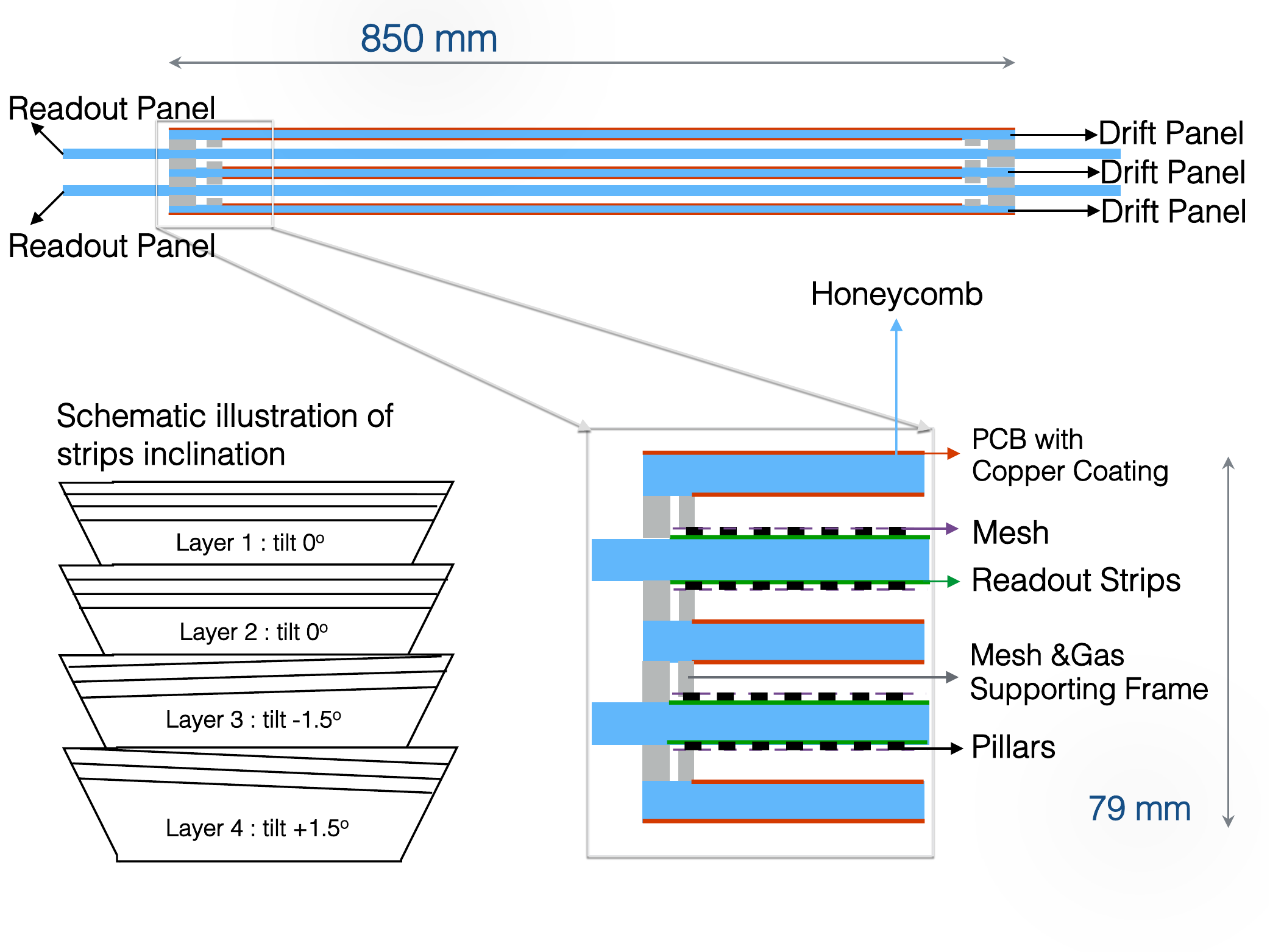}
	\caption{Layout of the MMSW detector; it consists of three drift and two readout panels (11.7~mm thick each), and four 5~mm thick gas gaps. In Layer 1 and 2 the readout strips are parallel, in Layer 3 and 4 the strips are inclined by $\pm$1.5$^\circ$.}
	\label{Fig:DetectorLayout}
\end{minipage}
\hspace{0.2cm}
\begin{minipage}[hbt]{.49\textwidth}
	\centering
	\includegraphics[width=0.98\textwidth]{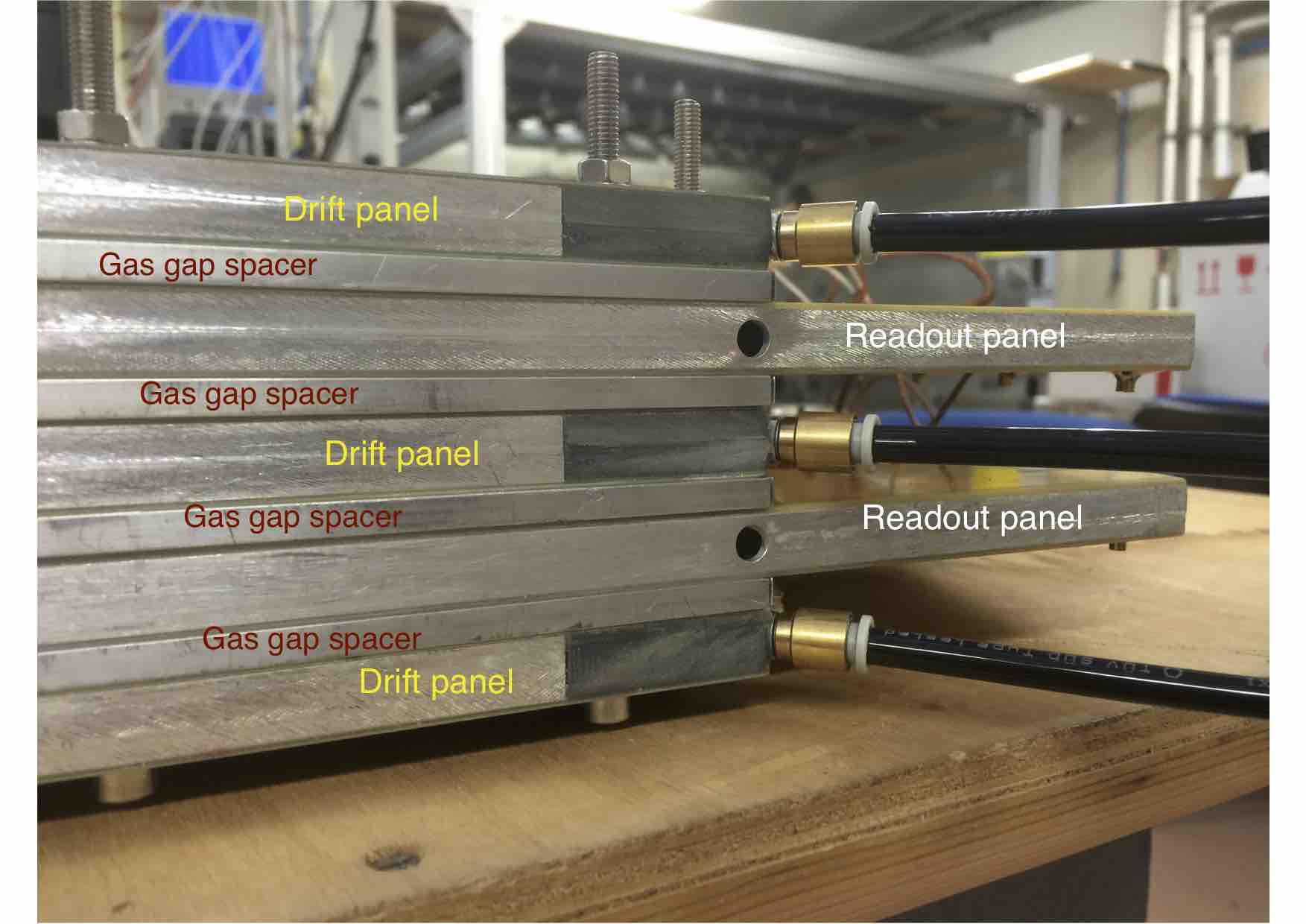}
	\caption{One corner of the micromegas quadruplet, showing the five panels, the gas gap spacers in between and the gas pipes.\vspace{1.3cm}}
	\label{Fig:GasGapSpace}
\end{minipage}
\end{figure}

Figure~\ref{Fig:DetectorLayout} shows a sketch of a MMSW quadruplet. The quadruplet consists of three drift panels, two readout panels and four gas gaps, created by spacer bars around the detector circumference. The panels are sandwiches consisting of two 0.5~mm thick skins (FR4 sheets) and a 10~mm thick stiffening element (Al frame + Al honeycomb). Details are given in Section~\ref{sec:Construction}. A photo of the assembly (Fig.~\ref{Fig:GasGapSpace}) shows the stack of the panels and spacer frames. 

The readout panels carry identical micromegas structures on both sides in a back-to-back arrangement. The micromegas structures are described in more detail in Sect.~\ref{sec:PCB}. They serve as skins of the panels and carry the readout strips that are running along the long dimension of the board. In the $\eta$ readout panel the strips on the two panel faces are parallel to each other. In the other panel the strips are inclined by $\pm$1.5$^\circ$ with respect to the direction of the parallel strips, forming a stereo pattern when mounted back-to-back. The stereo strips allow us to determine the position along the $\eta$ strip direction with a spatial resolution of 2--3~mm. The two stereo-strip layers together also measure the coordinate perpendicular to the strip direction with the same precision as the two parallel-strip layers. 

\begin{figure}[t]
    \begin{center}
        \includegraphics[width=0.55\textwidth]{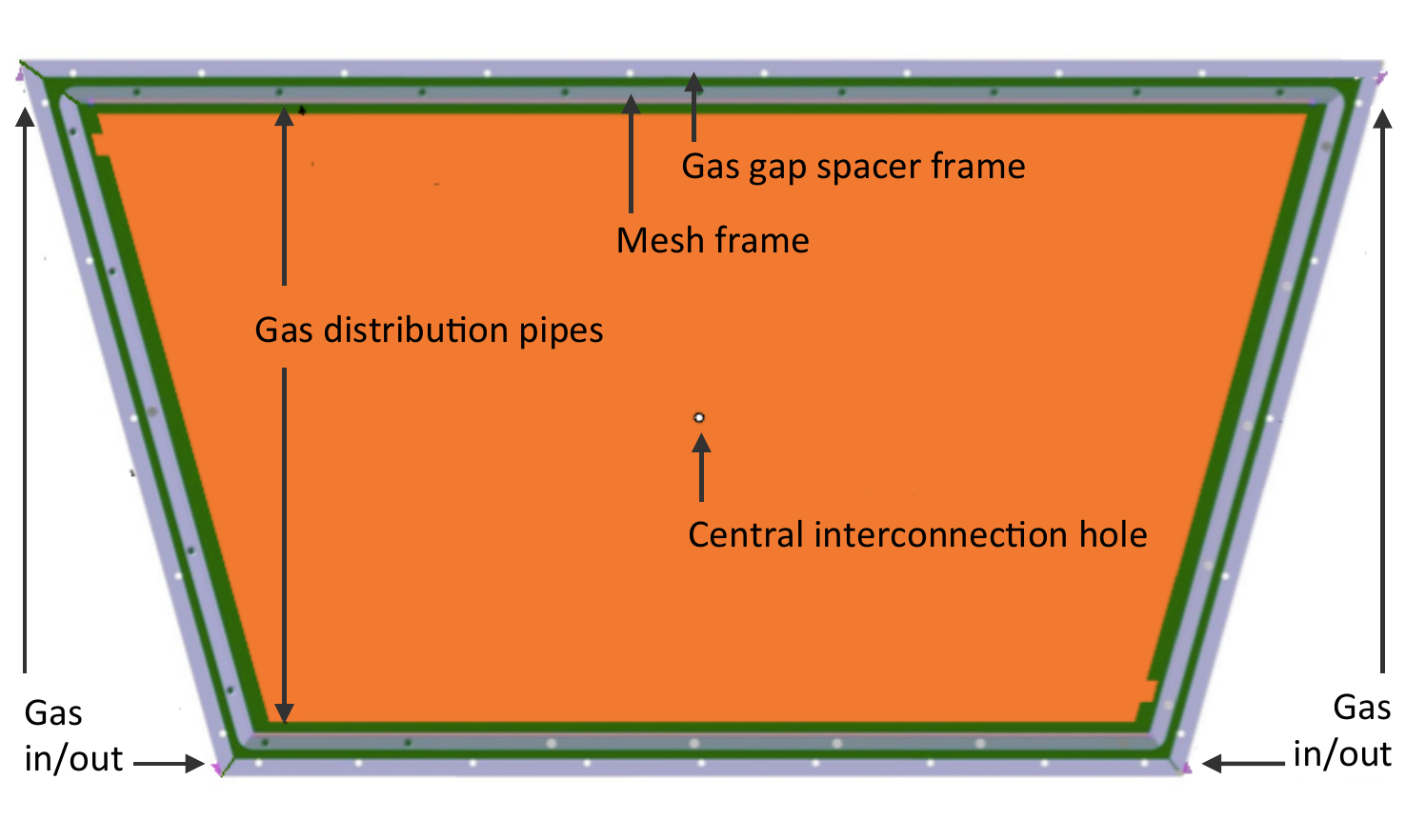}
        \includegraphics[width=0.42\textwidth]{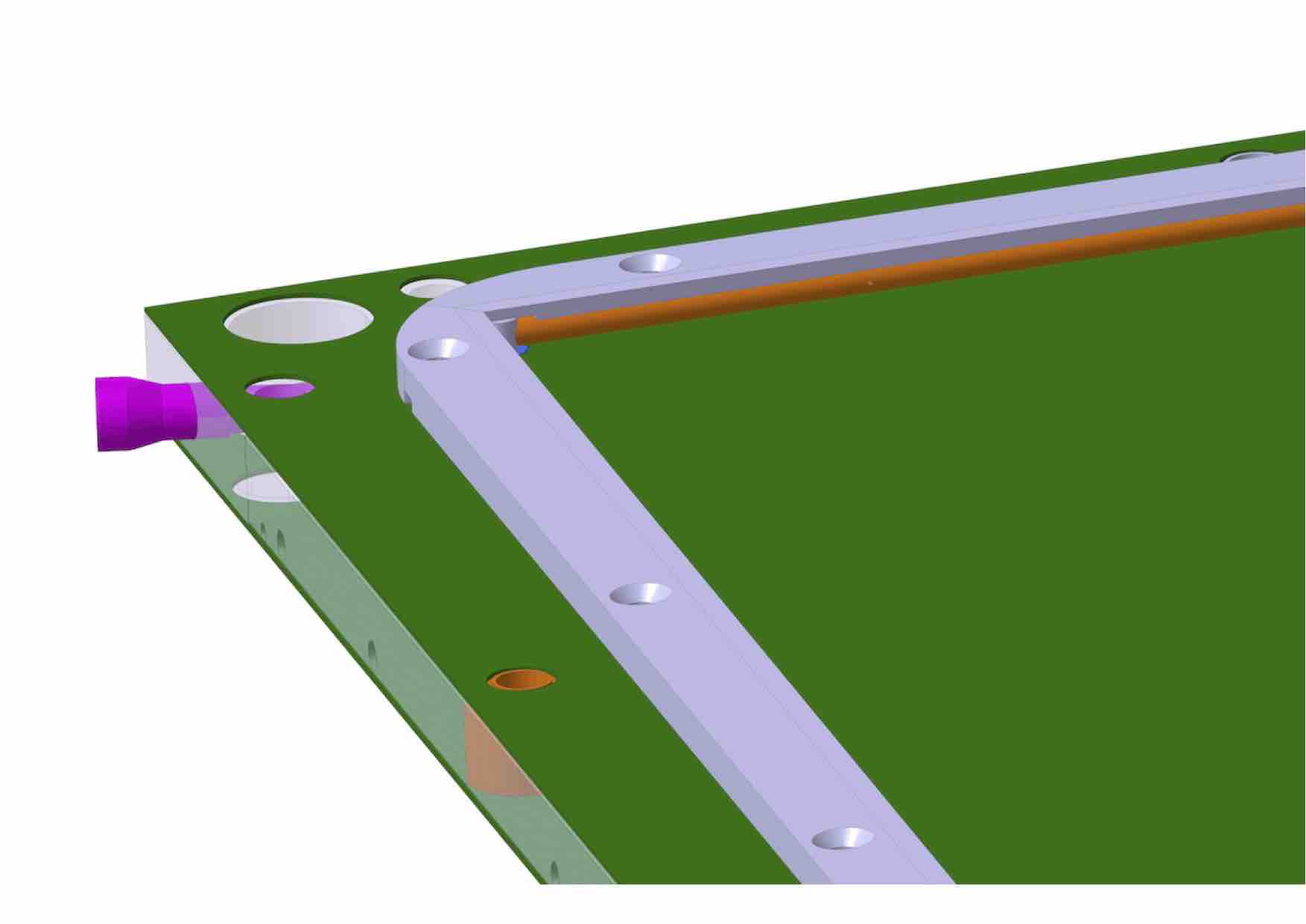}
        \caption{Left: PCB board for the drift panel, showing the drift electrode; right:  detail of the mesh frames with the assembly holes and the gas distribution feed-through and pipe.}
        \label{Fig:PCBDrift}
    \end{center}
\end{figure}

The drift panels integrate the Cu drift electrodes, the meshes, and the gas distribution system. Figure~\ref{Fig:PCBDrift} (left) shows the layout of the PCB  for the drift panels. In the two outer drift panels such PCBs constitute the skin facing the gas gap, in the central drift panel they are used for both skins. In the right illustration the corner of a drift panel with the mesh support frame mounted is shown.

Contrary to most micromegas detectors, we opted for a design where the mesh is not fixed to the readout board. In the following we call this design 'mechanically floating'\footnote{It should be noted that electrically the mesh is not floating but connected to the detector ground potential}. In view of the construction of 2--3~m$^2$ large detectors in which a single readout structure will consist of several PCBs, we chose to fix the mesh to the drift panel. In this design the mesh can stretch over several PCBs without creating dead areas. As a further benefit, the amplification gap is accessible and can be easily cleaned. 

\begin{figure}[htb]
\begin{minipage}[hbt]{.49\textwidth}
	\includegraphics[width=0.98\textwidth]{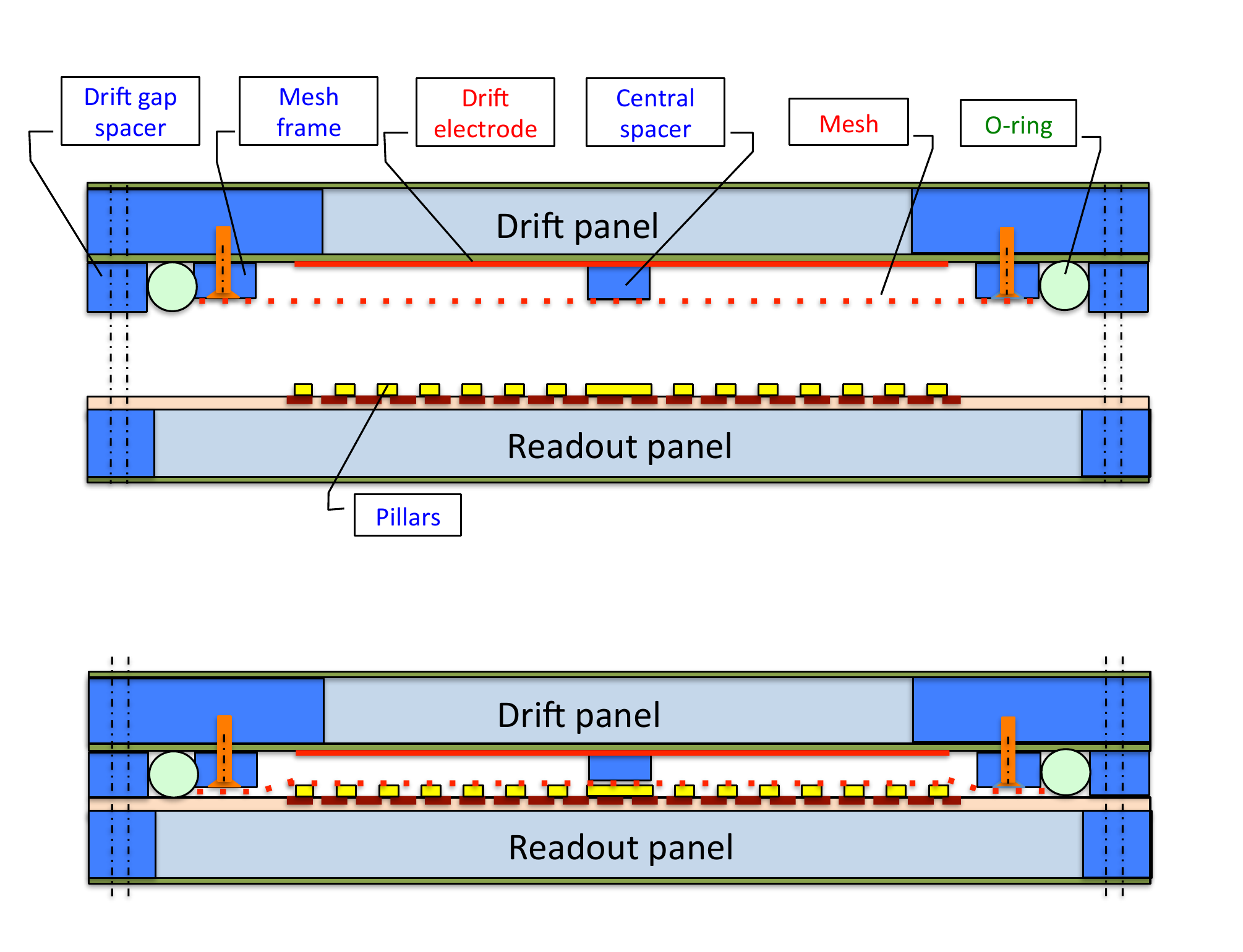}
	\caption{Illustration of the mechanically floating mesh concept showing the drift and readout panels in open and closed position. The mesh is fixed to the drift panel at a distance such that the mesh touches the pillars on the readout panel when the detector is closed (bottom figure).\vspace{1.8cm}}
	\label{Fig:MMSW_assembly_principle}
	\centering
\end{minipage}
\hspace{0.2cm}
\begin{minipage}[hbt]{0.49\textwidth}
	\centering
	\includegraphics[width=0.98\textwidth]{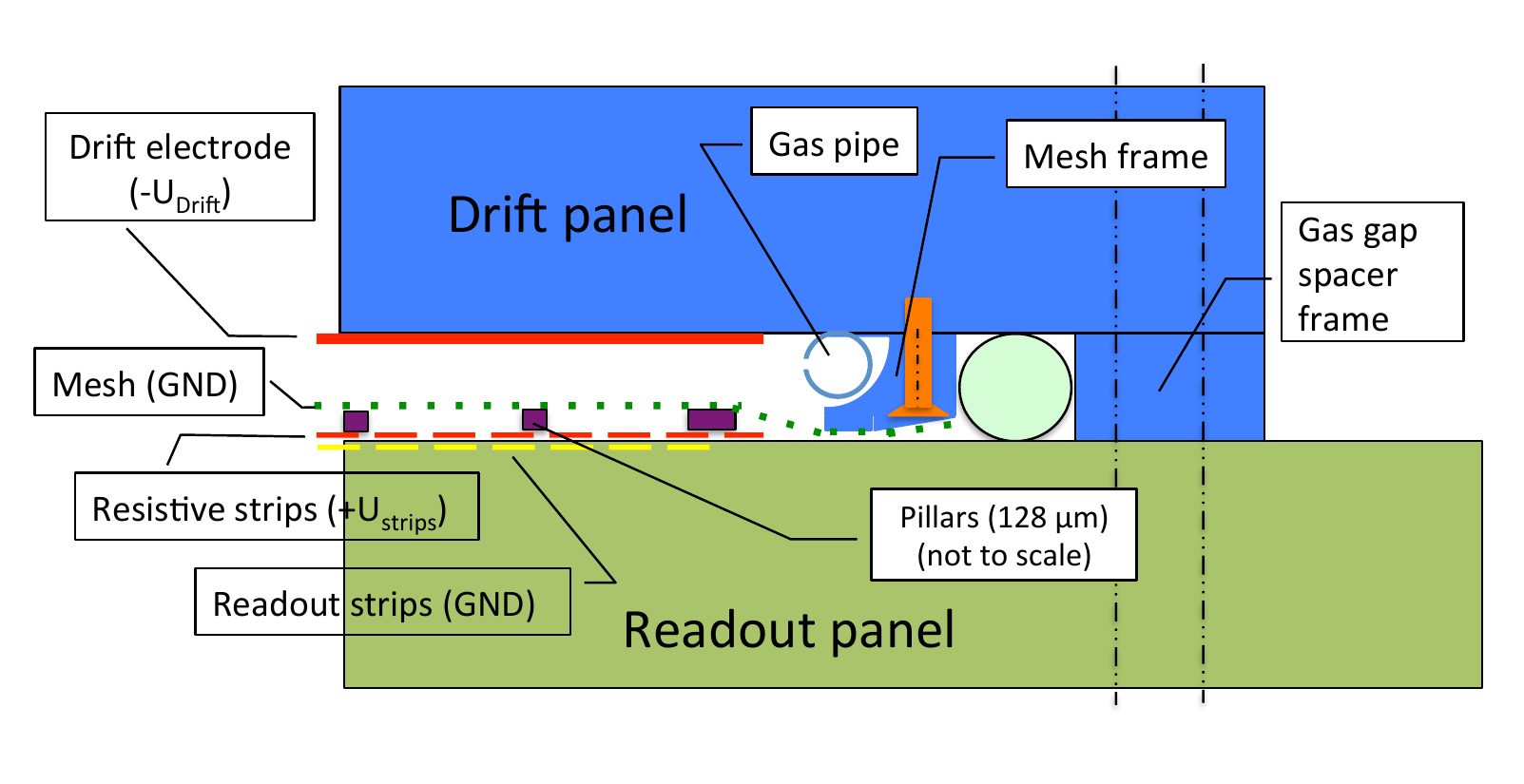}
	\caption{Schematics of the mesh and spacer frame arrangement (not to scale); the figure also indicates the electrical potentials of drift electrode, mesh, and resistive strips; the readout strips are, when connected to the readout electronics, virtually at ground (GND) potential.}
	\label{Fig:MMSW_stacking_detail}
\end{minipage}
\end{figure}

The mesh\footnote{For the MMSW detectors we use a woven stainless steel mesh with 30~$\mu$m wires and 50~$\mu$m opening.} is glued to the mesh frame. The latter is screwed and glued before to the drift panel. The mesh frame is $\sim$100~$\mu$m thinner than the spacers that define the gas gap thickness. When the readout panel, the gas gap spacers, and the drift panel are stacked, the mesh touches the pillars of the readout board. This is illustrated in Fig.~\ref{Fig:MMSW_assembly_principle} and Fig.~\ref{Fig:MMSW_stacking_detail}.

Electrically, the mesh is connected to the detector ground. The amplification field is created by polarising the resistive strips, see Sect.~\ref{sec:PCB}. The strong electrical field between the resistive strips and the mesh is not only used for the amplification process, the electrostatic force acting between it and the resistive strips also ensures that the mesh is attached to the pillars, defining an amplification gap of 128~$\mu$m.

Two 3~mm diameter stainless steel tubes, which are running on the inside of the mesh frame along the two long sides of the panel, see Fig.~\ref{Fig:PCBDrift} (left) and Fig.~\ref{Fig:MMSW_stacking_detail}, serve as gas distribution channels. The tubes have a number of small holes of different diameter along their length in order to equalise the gas flow. At both ends gas feed-throughs, made by 3D printing, connect the tubes to the external gas lines. The gas tightness is achieved by a 6~mm thick O-ring inserted between the mesh frame and the gas gap spacer, as illustrated in Fig.~\ref{Fig:MMSW_assembly_principle}.

The panels are held together by a series of bolts around the perimeter of the detector, approximately every 10~cm. Two of the assembly holes on the long side serve as precision alignment holes for the PCBs. They have tight-fit inserts (h6), one with a hole, the other with a slot. All other holes are non-precise and serve only for the detector assembly. In addition to the holes around the detector perimeter a 5~mm diameter hole in the geometrical centre of the board allows for the interconnection of the two outer panels. This interconnection corrects the small (O(0.5~mm)) deformation of the outer panels owing to the mesh tension. It also limits the deformation of the panels when the detector is operated with a gas overpressure of 1--2~mbar with respect to the atmospheric pressure. The drift and readout panels are 11.7~mm thick, each, the gas gap spacers are 5~mm thick, leading to a total thickness of the quadruplet of 79~mm, not counting the assembly bolts.

\subsection{Micromegas structure\label{sec:PCB}}

\begin{figure}[htb]
    \begin{center}
        \includegraphics[width=0.49\textwidth]{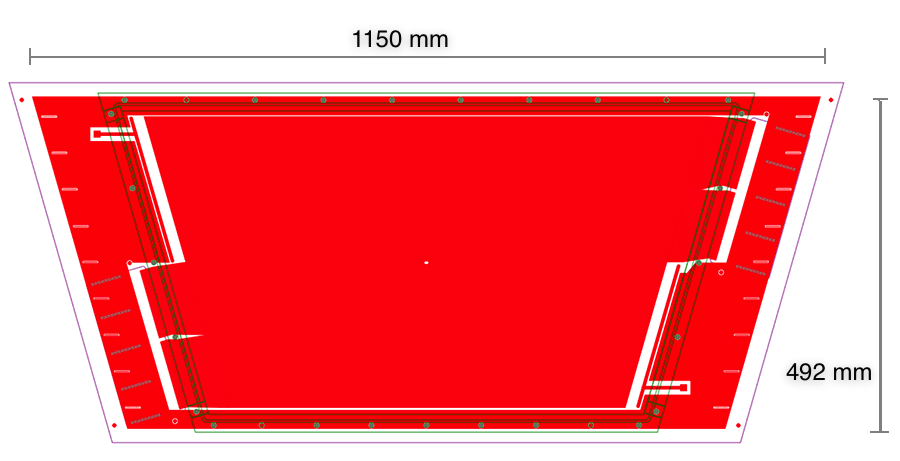}
        \includegraphics[width=0.49\textwidth]{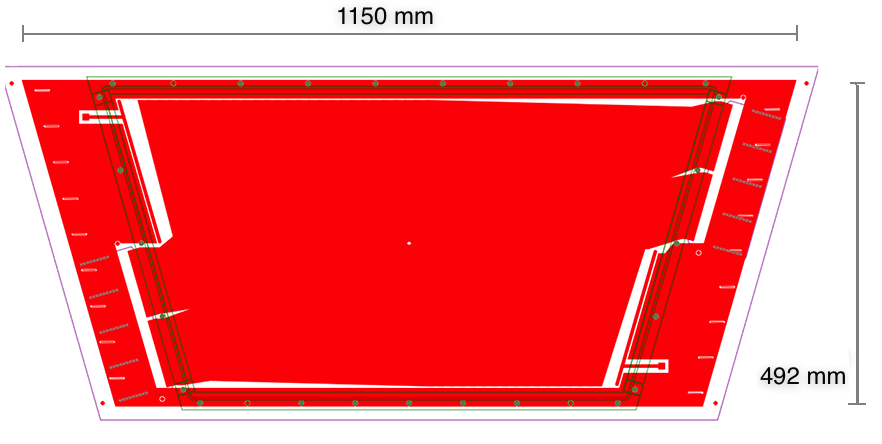}
        \caption{Layout of the PCB boards. For the $\eta$ board (left), the readout strips, the HV connection, the assembly holes, and the location of the readout electronics are shown, the short horizontal lines close to the inclined board edges give the position of the 1$^{st}$ and 128$^{th}$ strip, mod(128). In addition, the rotation of the readout strips by 1.5$^\circ$ is visible for the stereo board (right).}
        \label{Fig:ReadoutBoards}
    \end{center}
\end{figure}

The basic element of the micromegas structure is a 0.5~mm thick FR4 printed circuit board (PCB), clad with a 17~$\mu$m thick Cu layer on one side, the other side is bare FR4. Figure~\ref{Fig:ReadoutBoards} shows the layout of the $\eta$ and stereo readout boards. They have a trapezoidal shape, with an upper base of 1150~mm, a lower base of 870~mm, and a height of 492~mm. Each PCB comprises 1024 readout strips with lengths increasing from $\simeq$600~mm to $\simeq$800~mm between the short and long sides of the boards. The strips have a width of 300~$\mu$m and a pitch of 415~$\mu$m. 
 
The strip pattern on the boards with the stereo strips is identical to the one with the $\eta$ strips. The latter is simply rotated by 1.5$^\circ$ around the central hole position. By rotating the strip pattern part of the first and last 20--24 strips at the upper and lower edges move out of the area defined by the $\eta$ strips. These strips were shortened accordingly. The rotation leads, in addition, to two triangular regions on the upper and lower edges that are not covered by readout strips.  
 
The readout strips are split into two groups. The upper 512 strips are routed out of the active area to the right side of the board, the lower 512 strips to the left side. They end approximately 10~mm outside the detector frame. Each group of 512 strips is again split into two groups of 256 strips that are routed out between the assembly holes to the readout contacts. The area on the sides of the boards accommodates the readout electronics. Given the large number of readout strips in the NSW system, we use the MMSW detector to test the Zebra\footnote{see e.g., ZEBRA$^{\textregistered}$ connectors from Fujipoly$^{\textregistered}$: http://www.fujipoly.com/usa/products/zebra-elastomeric-connectors/} connection system that avoids soldering on the PCBs. 

\begin{figure}[htb]
    \begin{center}
        \includegraphics[width=0.95\textwidth]{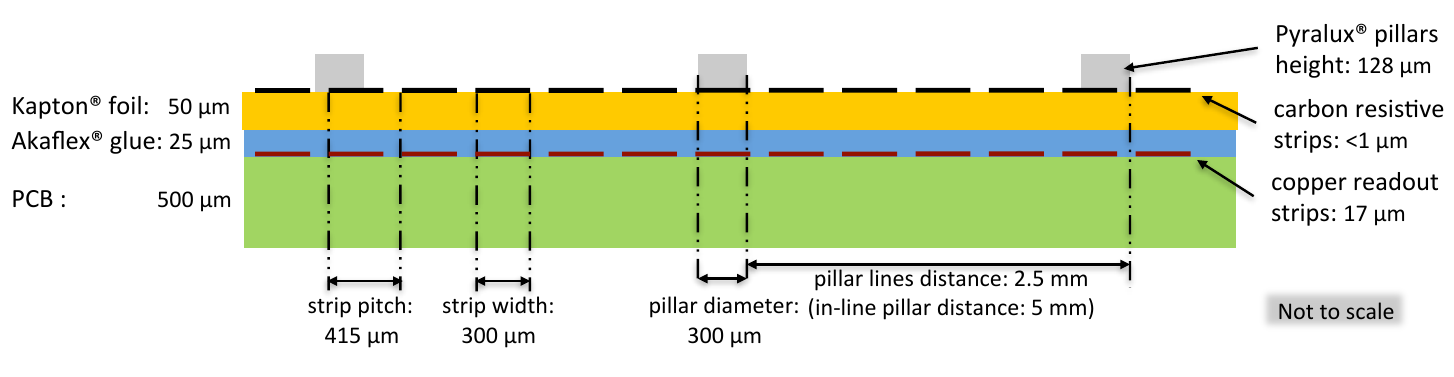}
        \caption{Cross section through the micrometers structure (schematic) showing the arrangement and dimensions of the elements of the structure.}
        \label{Fig:MMstructure}
    \end{center}
\end{figure}

The micromegas structure is built on top of the PCB, as illustrated in Fig.~\ref{Fig:MMstructure}. 

The readout strips are covered by a 50~$\mu$m thick Kapton$^{\textregistered}$  foil comprising carbon resistive strips. The resistive strips were produced in Japan by sputtering \cite{Ochi:2014rda} and have a resistivity of 0.5--1~M$\Omega/\Box$. The foil is glued on top of the readout PCB. The resistive strips serve as protection to minimise the effect of sparks~\cite{Alexopoulos:2011zz} by limiting the spark currents. Signals are induced via capacitive coupling to the readout strips. 

\begin{figure}[htb]
\begin{minipage}[hbt]{0.39\textwidth}
	\centering
	\includegraphics[width=0.98\textwidth]{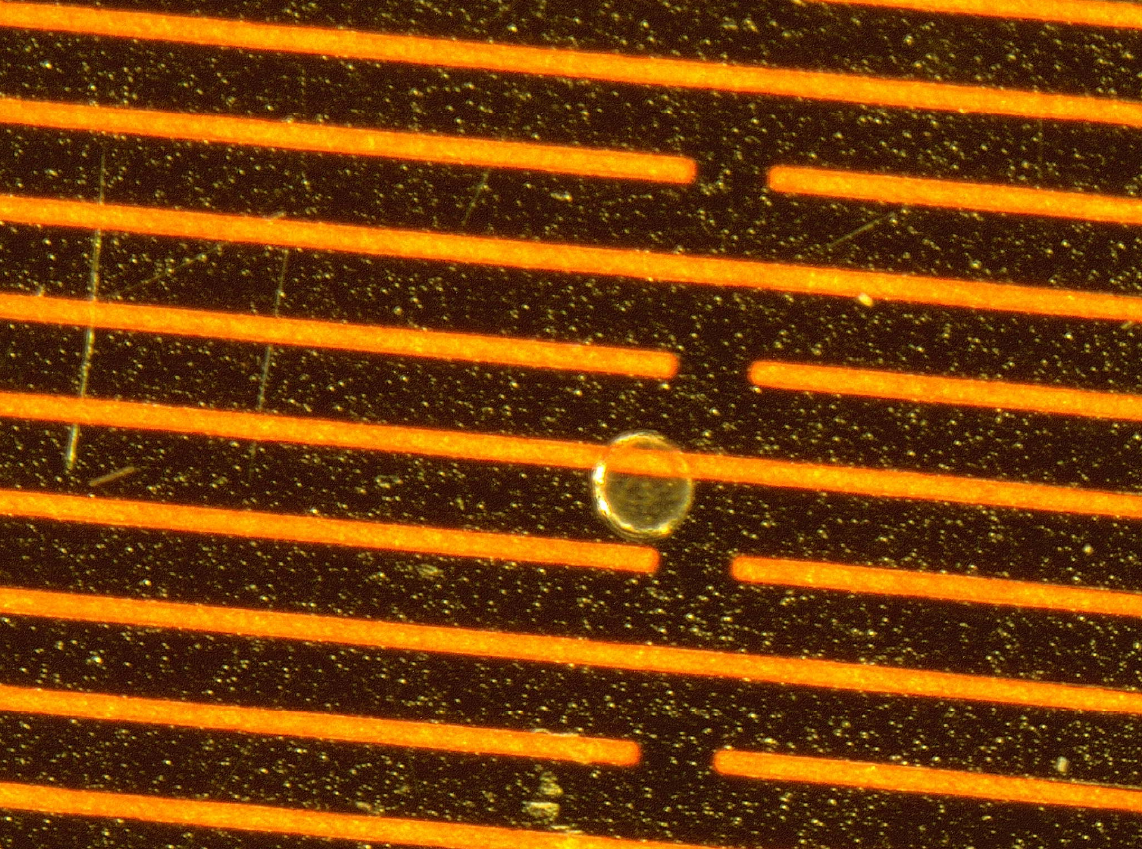}
	\caption{Photo of the resistive strips. The resistive strips with their interconnects are dark coloured; also visible are parts of the readout strips in the layer below (bright) and one pillar.}
	\label{Fig:Resistive_strips}
\end{minipage}
\hspace{0.2cm}
\begin{minipage}[hbt]{.59\textwidth}
	\centering
	\includegraphics[width=0.95\textwidth]{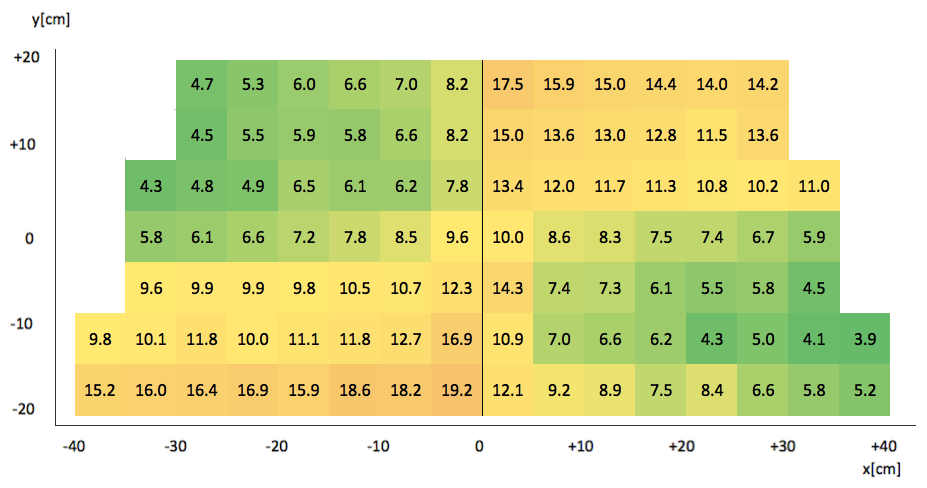}
        \caption{Mapping of the resistivity on one readout layer. The impedance is measured between the HV supply point, on the top-left side or bottom-right side for the two sectors, and the indicated spot on the layer, resulting in a systematic asymmetry. \vspace{0.2cm}}
        \label{FIG:RESISTIVITY}
\end{minipage}
\end{figure}

The resistive strips follow the same overall pattern as the readout strips, with a few differences. The resistive strips are split in the middle. Each side is (or can be) connected to its own HV potential. In addition, as can be seen in the photo in Fig.~\ref{Fig:Resistive_strips}, neighbouring strips are interconnected between each other every 20~mm along the strips. The connection bridge is shifted by 10~mm from one strip to the next.  Thanks to these interconnections the charge is evacuated not only through a single strip but through a network of strips thus equalising the effective resistances over the full detector area\footnote{The local resistance (the charge along the resistive strips spreads over a few mm) is not much affected by these interconnects} to $5-10$~M$\Omega$, see Fig. \ref{FIG:RESISTIVITY}. Another asset of this scheme is that defects in the resistive-strip pattern, e.g., broken strips, become uncritical (unless too massive).

On top of the Kapton foil with the resistive strips a pattern of of 128~$\mu$m high pillars with a diameter of 300~$\mu$m has been created using photolithography. The pillars are placed at a distance of 5~mm of each other along the direction of the readout strips. The next row of pillars is at a distance of 2.5~mm, however, the pillars are shifted by 2.5~mm along the strip direction. In Fig. \ref{Fig:Resistive_strips} a pillar is also visible.

\section{\label{sec:Construction}Detector construction}

In the following, we describe the construction of the MMSW detectors. Several improvements have been implemented in the construction of the second detector (MMSW-2), based on the experience gained during the construction of the first quadruplet (MMSW-1). The following description focusses on the technique used for the construction of MMSW-2.

\subsection{Construction of the vacuum tables\label{sec:VacuumTable}}

\begin{figure}[t]
    \begin{center}
        \includegraphics[width=0.49\textwidth]{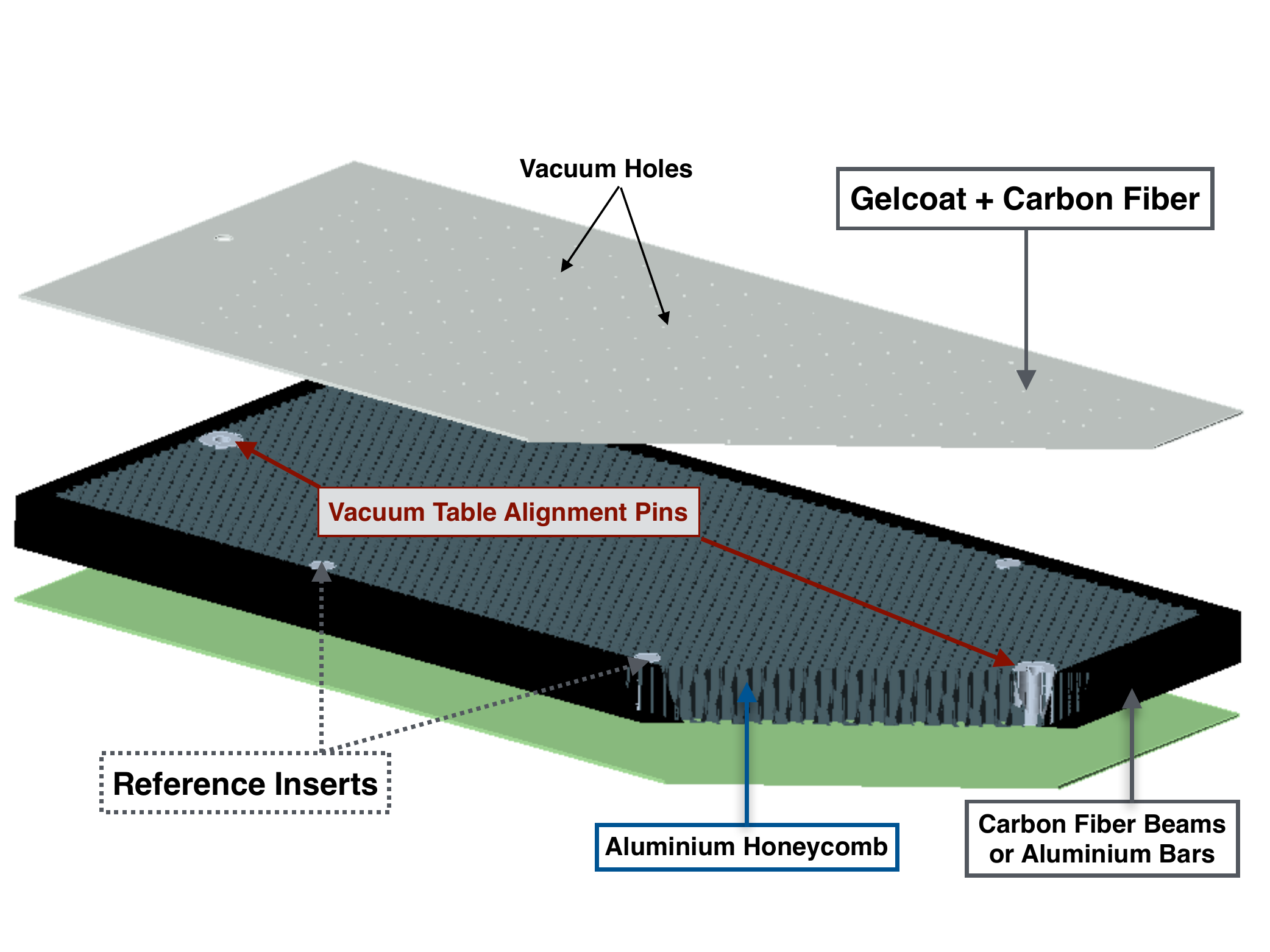}
        \includegraphics[width=0.49\textwidth]{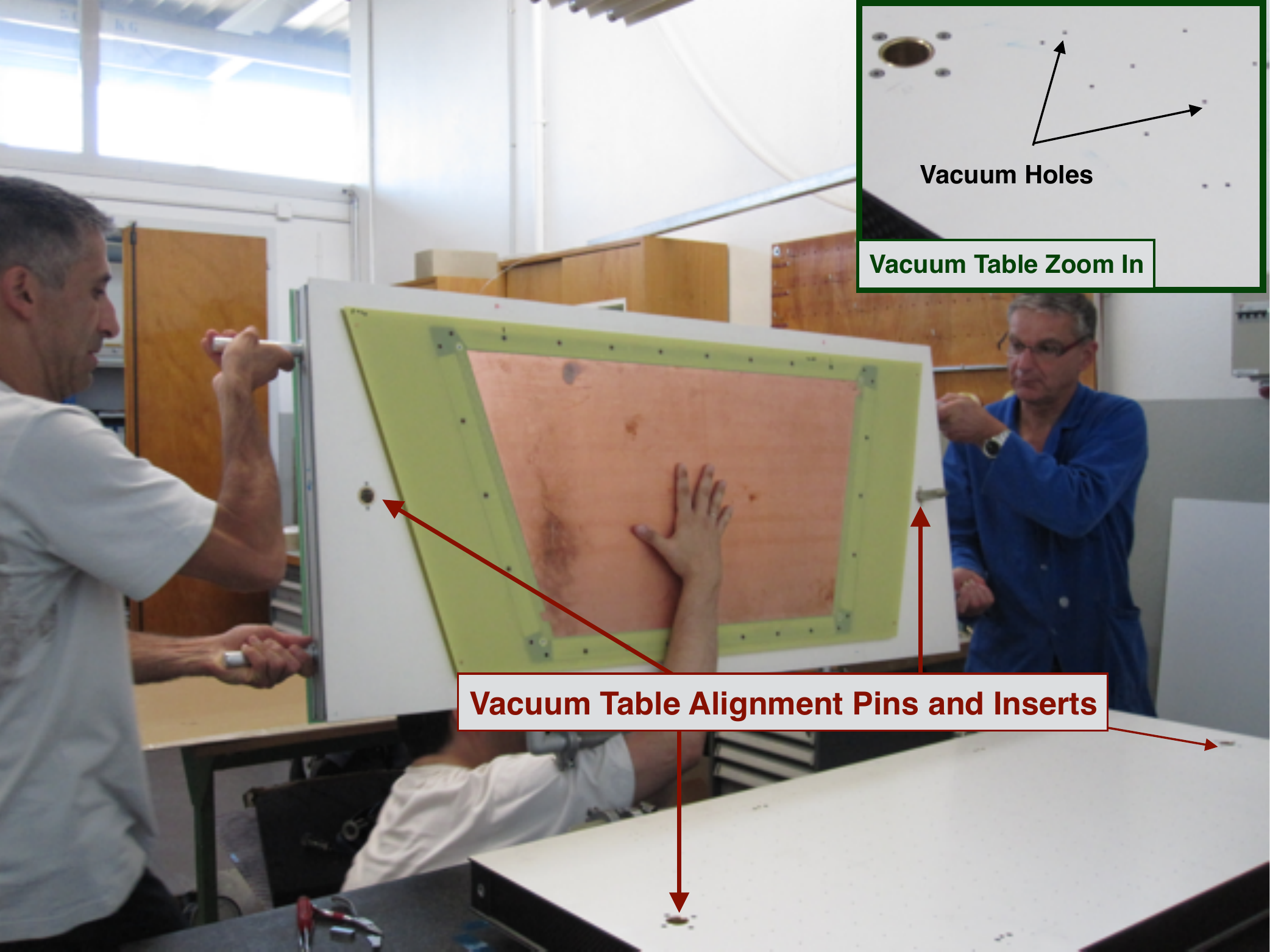}
        \caption{Schematic sketch of the vacuum table (left) and photo of the two stiff-backs (right)}
        \label{Fig:Vacuum}
    \end{center}
\end{figure}

In order to reach the required spatial resolution of the detector, a good planarity of the readout and drift panels is required. This is achieved by two vacuum tables, called 'stiff-backs' in the following. The stiff-backs are made of 60 mm perforated aluminium honeycomb material sandwiched between two 3~mm thick skins of glass fibre and aluminium profiles as frame, see Fig.~\ref{Fig:Vacuum} (left). The honeycomb provides for the stiffness of the table structure. One of the glass fibre surfaces is coated by a 0.5~mm thick gel-coat layer serving as precision surface. The glass fibre skin together with the gel-coat layer have been prepared and cured on a high-precision granite table transferring the granite table flatness to the stiff-back. 

In a second step, sucking holes with a diameter of 1\,mm and a distance of 50\,mm are drilled into the gel-coat cover. The perforated honeycomb allows for a uniform distribution of the vacuum across the full surface. In addition to the sucking holes, two large-diameter inserts for the stiff-back alignment and two small-diameter inserts for the panel positioning on the stiff-back are implemented. These can be seen in the photo (Fig.~\ref{Fig:Vacuum} (right)) taken during tests prior to the glueing of the panels. It shows the two stiff-backs with a drift-electrode PCB attached to the upper stiff-back.
 
\begin{figure}[t]
    \begin{center}
        \includegraphics[width=0.49\textwidth]{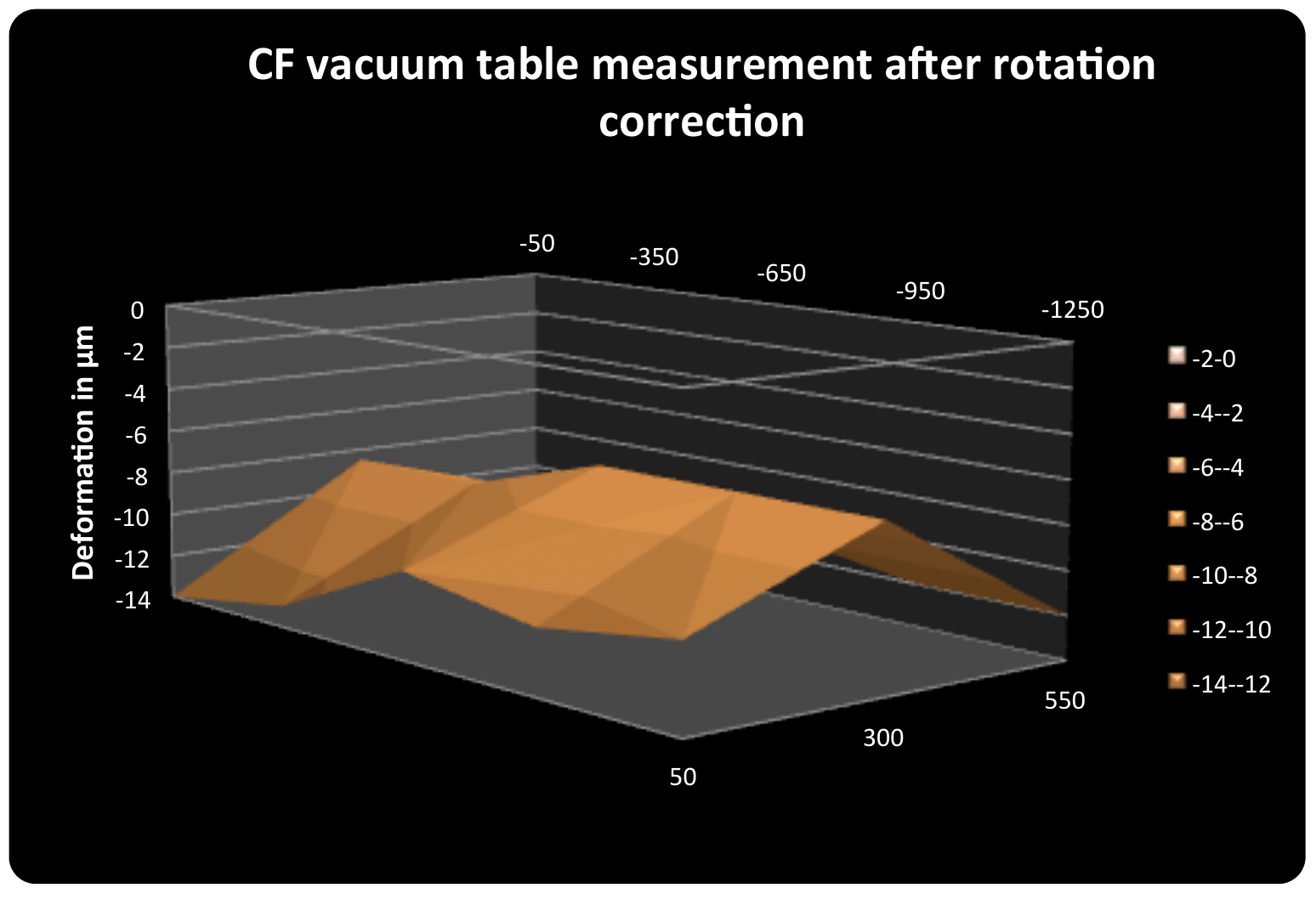}
        \caption{Flatness measurement of the carbon fibre table.}
        \label{Fig:C-fibrePlanarity}
    \end{center}
\end{figure}

During the glueing of the panels for MMSW-1, a temperature dependance of the stiff-back flatness was observed. Therefore, a new stiff-back was built with carbon-fibre\footnote{Carbon-fibre has a significantly smaller temperature expansion coefficient than glass fibre and aluminium.} as skin and frame material, however, still using the same aluminium honeycomb material. The planarity of the carbon-fibre stiff-back is shown in Fig.~\ref{Fig:C-fibrePlanarity}, similar to the one obtained for the Al/glass-fibre stiff-backs, however, constant with temperature.

Only one carbon-fibre table was built. It was employed as upper stiff-back in the construction of the MMSW-2 panels and the refurbishment of the MMSW-1. The bottom aluminium table, with its shortfalls, was used throughout all panel glueings.

\subsection{Panel glueing\label{sec:Panel_glueing}}

Each panel consists of two 0.5~mm thick FR4 skins, glued onto a 10~mm thick sheet of perforated aluminium honeycomb material \footnote{The honeycomb cells are 9~mm wide.} and a 10~mm high and 30~mm wide aluminium profile with 1.5~mm wall thickness. 

\begin{figure}[htb]
    \begin{center}
        \includegraphics[width=0.99\textwidth]{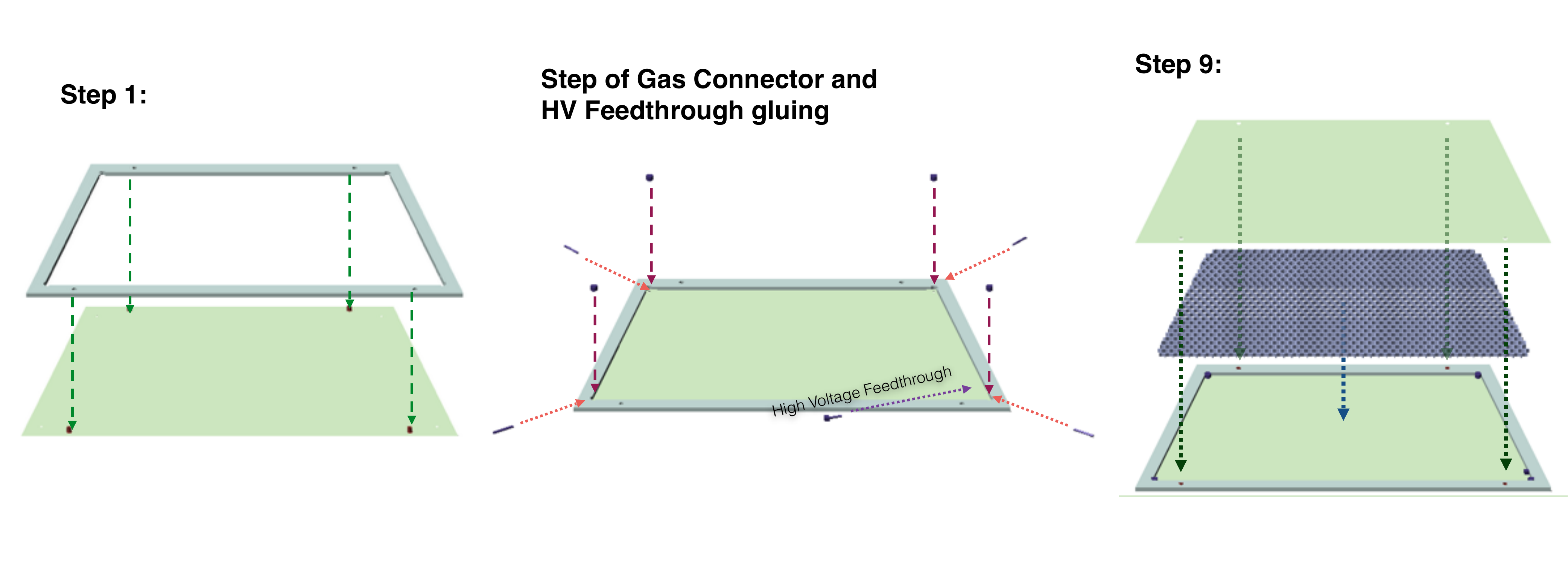}
        \caption{Schematic illustration of the drift panel construction. }
        \label{Fig:DriftSchematic}
    \end{center}
\end{figure}

The panel construction is illustrated in Fig.~\ref{Fig:DriftSchematic} and described in the following. It is to a large extend identical for the drift and readout panels, with a few exceptions that will be mentioned below.

\begin{figure}[htb]
    \begin{center}
	\includegraphics[width=0.295\textwidth]{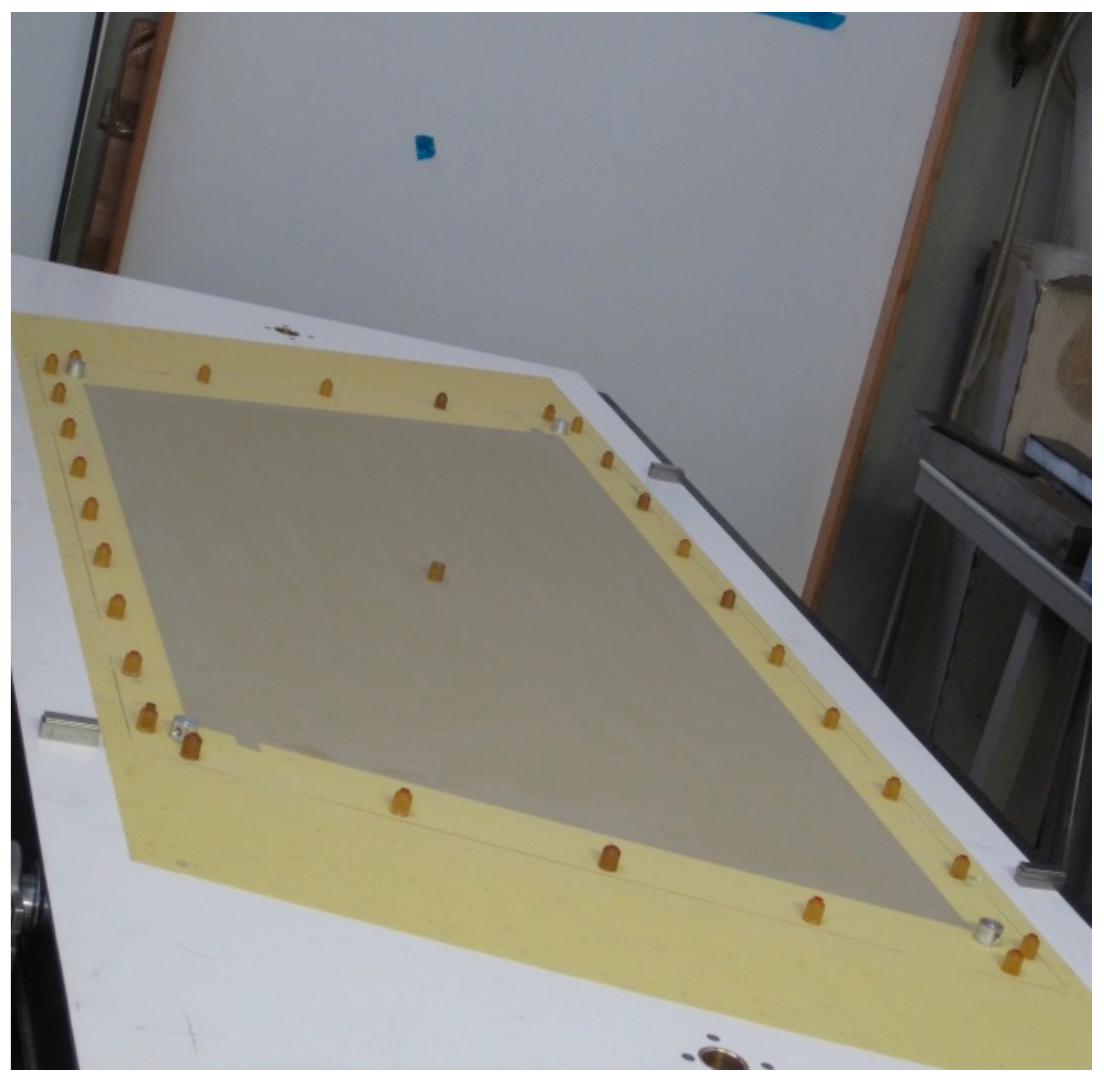}
	\includegraphics[width=0.295\textwidth]{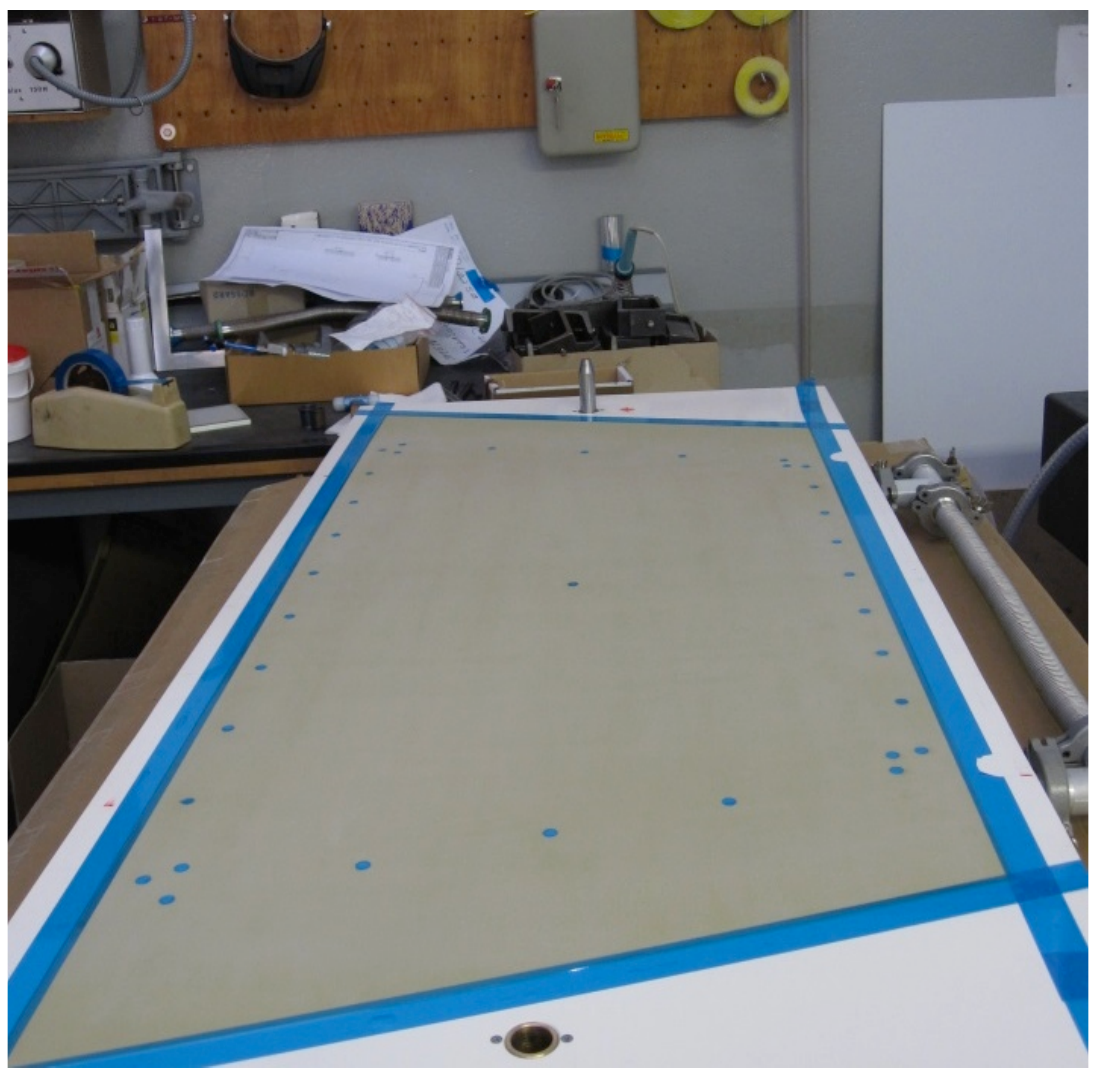}
	\includegraphics[width=0.38\textwidth]{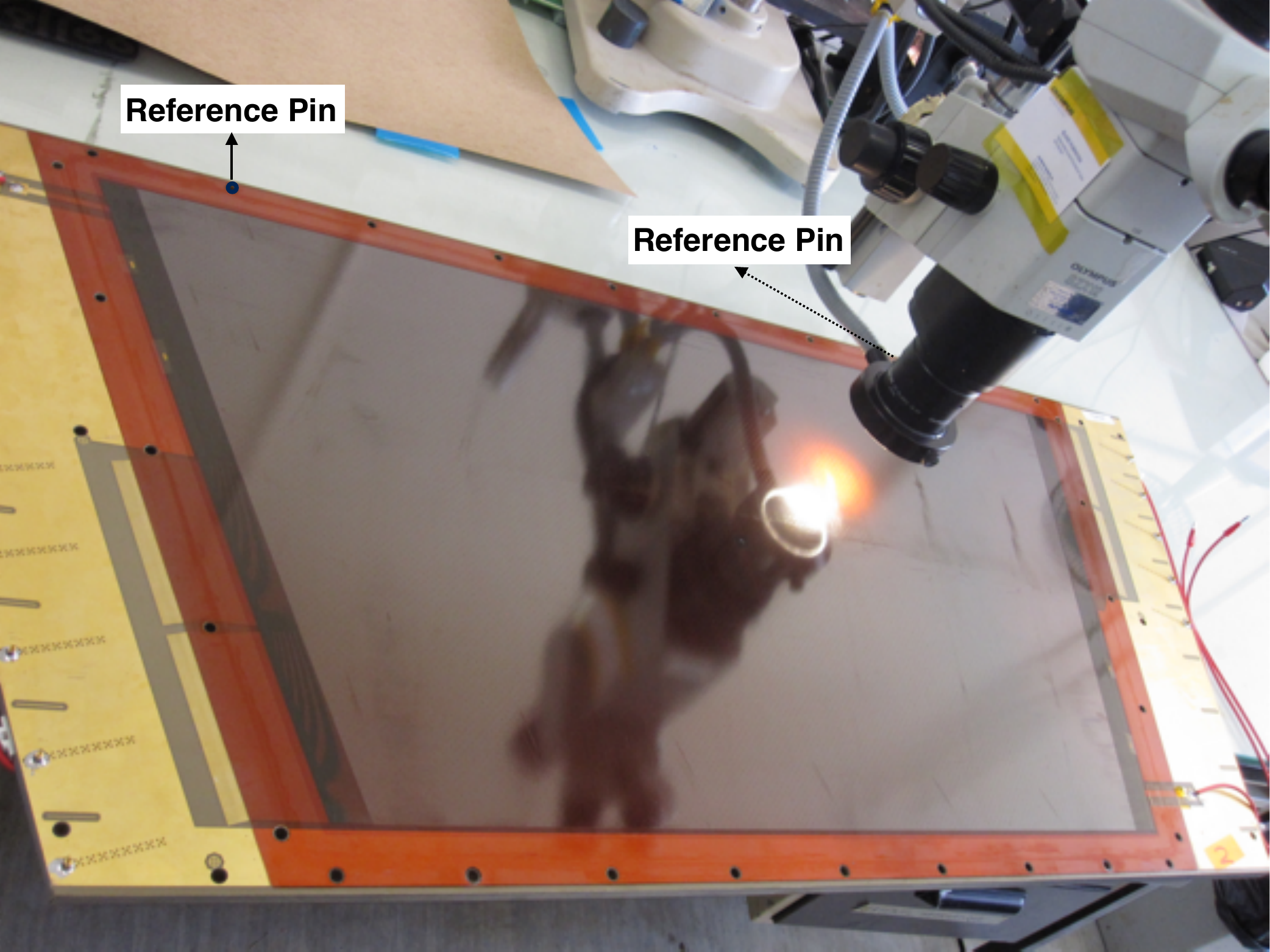}
        \caption{Left and middle: Skins on the two stiff-backs before glueing; right: readout panel with reference pins.}
        \label{Fig:skins_on_stiff-back}
    \end{center}
\end{figure}

\begin{figure}[htb]
    \begin{center}
        \includegraphics[width=0.49\textwidth]{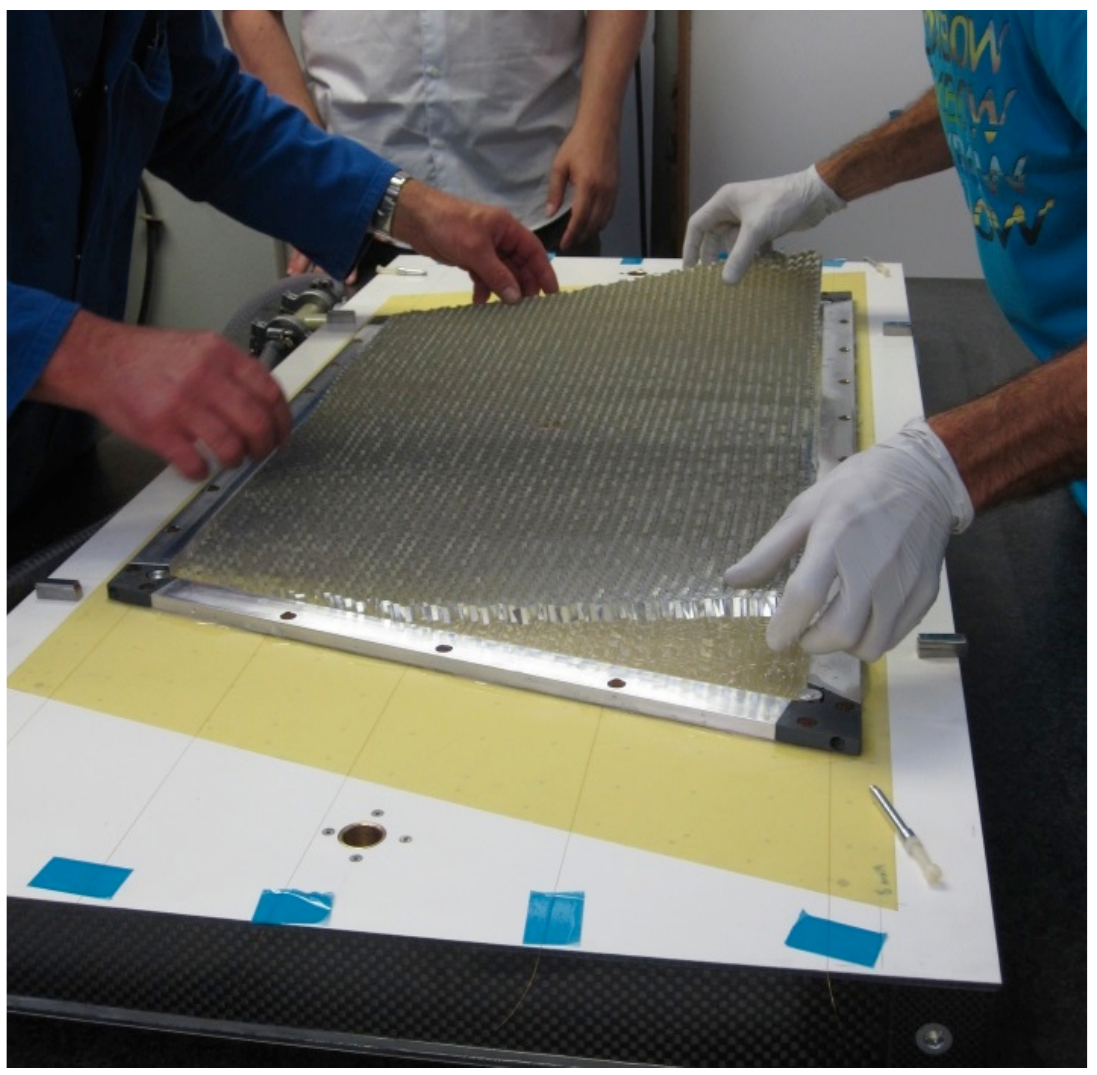}
        \includegraphics[width=0.49\textwidth]{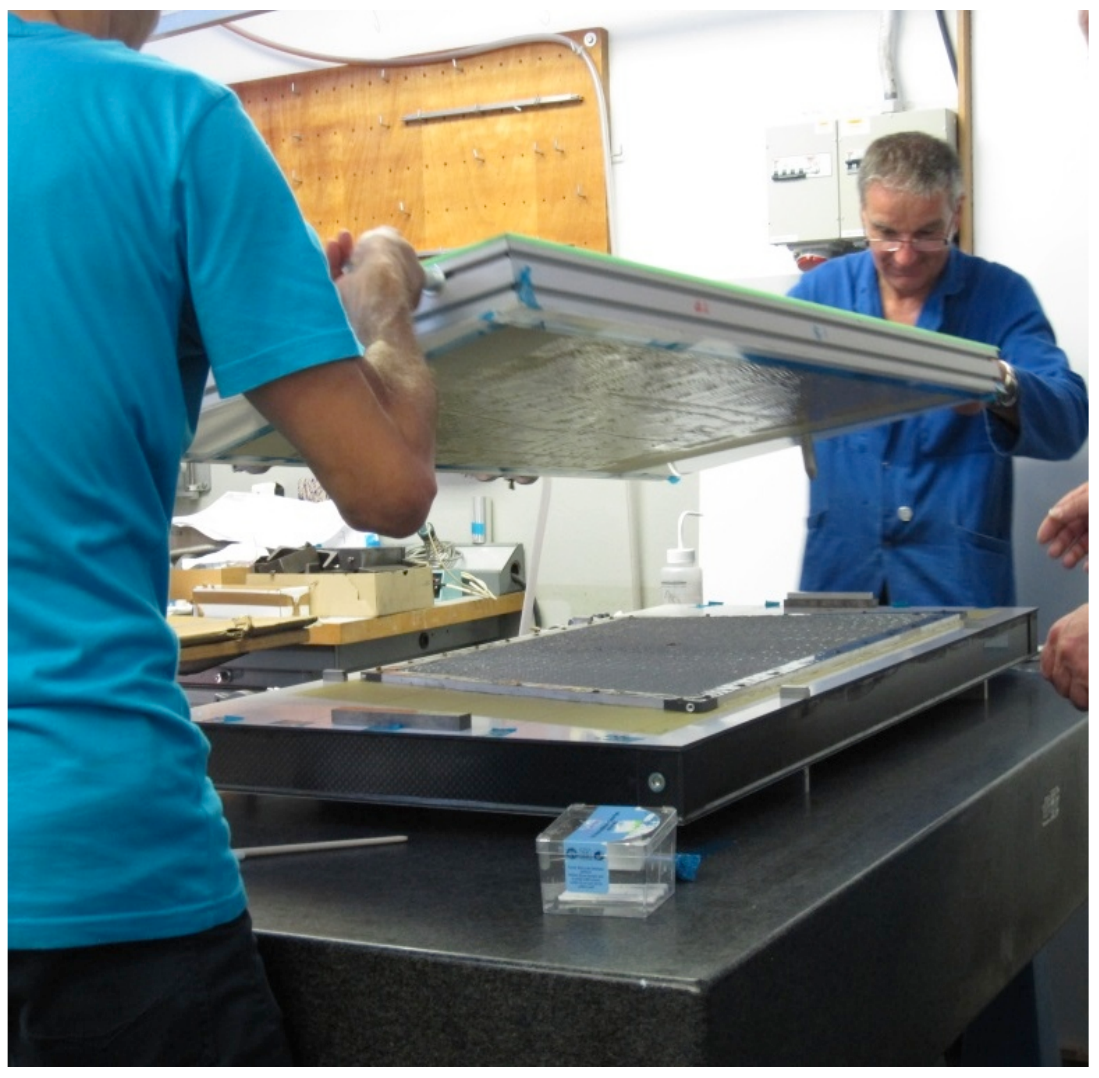}
        \caption{Frames and the honeycomb sheet are placed on the lower skin (left) and lowering of the upper stiff-back for the final sandwhich construction (right).}
        \label{Fig:Panel_glueing}
    \end{center}
\end{figure}

\begin{enumerate} 
\item Peek inserts are glued into the assembly holes of one of the two FR4 skins of a panel (bare PCBs for the drift panel, the micromegas structure with the pillars for the readout panel). A template guarantees the exact positioning of the peek inserts. For most of the peek inserts no particular precision is required. Two of the peek inserts along the long side of the PCBs, however, are precision parts and serve for the alignment of the two skins. 
\item The FR4 skin is placed with the copper-coated face (or the Micromegas structure) down and the peek insert facing up on the lower stiff-back, as shown in Fig.~\ref{Fig:skins_on_stiff-back} (left); it is positioned by pins in the precise alignment holes.
\item The skin is sucked on the assembly table by the vacuum pump, the edges are sealed with adhesive tape. This is shown in Fig.~\ref{Fig:skins_on_stiff-back} (middle) for the other skin.
\item A glue layer is distributed on the FR4 surface.
\item Five 250~$\mu$m thick stainless steel  wires are stretched over the FR4 sheet, they define the glue gap height, see Fig.~\ref{Fig:Panel_glueing} (left).
\item The frames and the honeycomb sheet are placed on the lower skin; the positioning of the frames is guaranteed by the peek inserts
\item The second FR4 skin is placed with the Cu electrode or the micromegas structure facing down on the upper stiff-back and sucked, see Fig.~\ref{Fig:skins_on_stiff-back} (middle). 
\item A uniform glue layer is distributed on the second surface similar to Step 4
\item The upper stiff-back, with the second skin attached, is placed above the lower stiff back and lowered onto a set of precise shims, see Fig.~\ref{Fig:Panel_glueing} (right). The shims define the glue gap between the upper skin and the frame and honeycomb as well as the overall thickness of the panel. The alignment of the two stiff-backs is obtained with the help of reference pins inserted into the large precision inserts that are visible in Fig.~\ref{Fig:Panel_glueing} (left). 
\item The glue is cured for 24 hours.
\end{enumerate} 

The main difference in the glueing of the readout panels with respect to the drift panels comes from the different external surface structures of the skins. The PCB surfaces of the drift panels are either flat or contain only a 17~$\mu$m thick Cu layer. The micromegas structures, discussed in Sect.~\ref{sec:PCB}, that serve as skins of the readout panels have thickness differences of up to 180~$\mu$m owing to the resistive foil layer and pillars that do not extend over the full PCB surface. Wherever necessary, three layers of 60~$\mu$m thick adhesive tape are glued to the PCB to compensate for the thickness differences. This is done before placing the micromegas structure on the stiff-back and avoids the deformation of the skin when the vacuum sucking is applied.

Glueing the two skins simultaneously to the Al-honeycomb structure saves time, however, introduces a small asymmetry in the glue distribution around the honeycomb cells on top and bottom. 

\subsection{Panel services and mesh glueing \label{sec:Panel_services}}

While little extra work after panel glueing is needed for the readout panels, the drift panels need to be equipped with the mesh frame, the gas distribution, high voltage (HV) connection, and the mesh, in this order. 

\subsubsection{Mesh frame\label{sec:Mesh_frame}}

As explained in Sect.~\ref{sec:Layout}, the mesh is connected to the drift-panels. Four straight bars are glued with Araldite 2011 to the drift panels, about twenty millimetres inside from the panel edges, serve as mesh frame. The bars are 16~mm wide and have a chamfer with an angle of about 3$^\circ$ on the outer part of the upper surface, leaving a 4~mm wide flat area. The two bars on the long sides contain a groove in their lower part to host the gas distribution pipe. Openings in the mesh frame at each corner of the panel serve as evacuation paths for the water during later cleaning of the panels. 
Figure \ref{Fig:MeshFrame} shows a photo of the mesh frame mounted to the drift panel.

After mounting the mesh frame, an insert is glued on the panel around the central hole, where the screw for interconnecting all the panels of the detector will be inserted. The central insert is made of peek and has the same height as the mesh frame. At its upper surface it contains a circular groove to absorb possible excess glue during the mesh gluing. 

This step is followed by a panel leak test, closing the panel with a temporary cover that is sealed with an O-ring around the mesh frame.

\subsubsection{Gas and HV distribution\label{sec:Gas_distribution}}

The gas is injected through the drift panel. For this purpose inserts in the corners of the drift panel frame serve as gas feed-throughs. These parts have been produced by 3D printing and are part of the panel frame. 

Before glueing the mesh, the gas distribution pipes are mounted. They are held in place by junction pieces that are glued at both sides to the feed-throughs and the gas pipes. 

\begin{figure}[htb]
\begin{minipage}[hbt]{0.45\textwidth}
	\centering
	\includegraphics[width=0.98\textwidth]{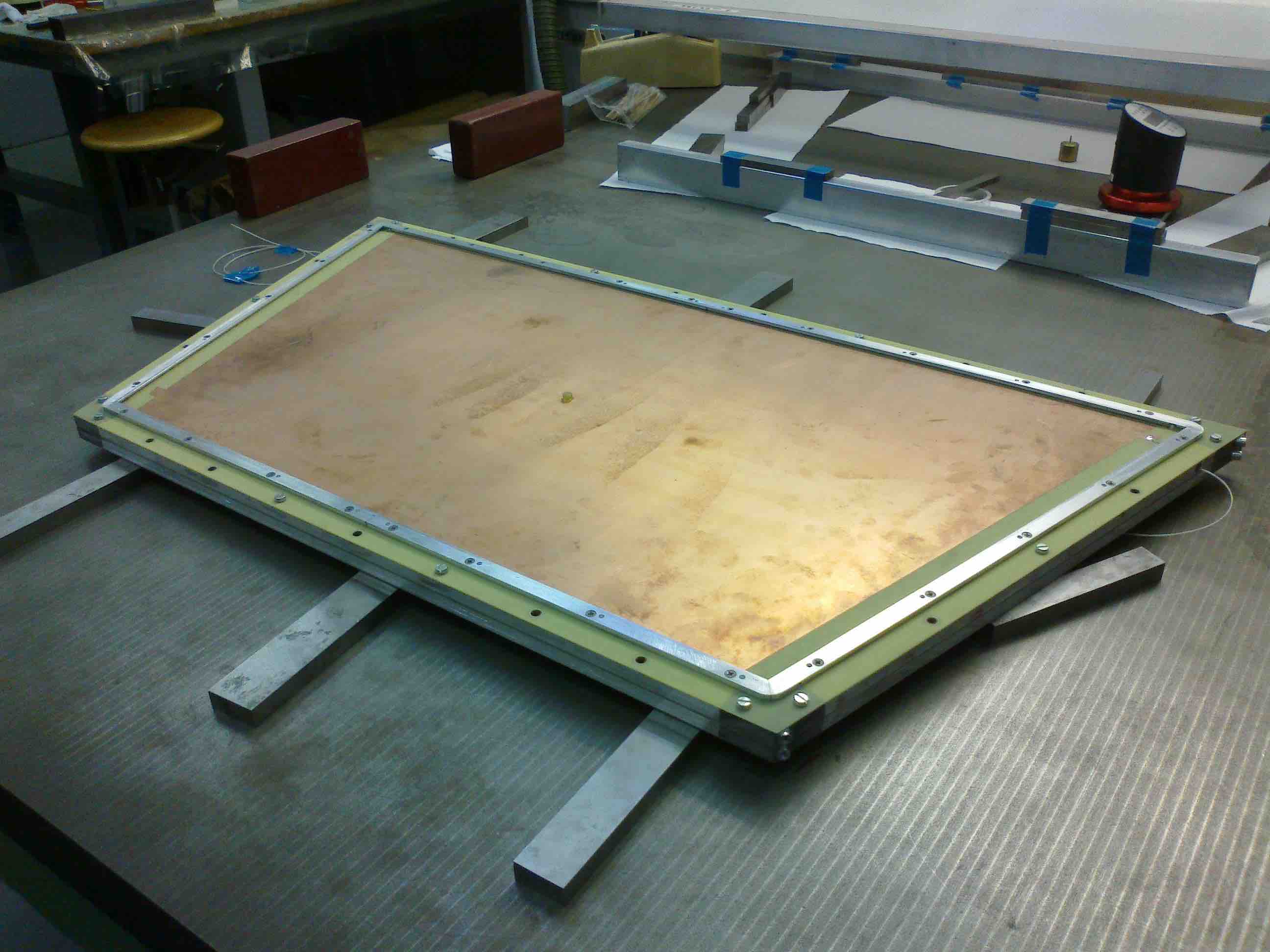}
	\caption{Mesh frame glued to the drift panel, in the centre of the drift electrode the small insert (see text) is also visible.\vspace{0.7cm}}
	\label{Fig:MeshFrame}
\end{minipage}
\hspace{0.2cm}
\begin{minipage}[hbt]{.54\textwidth}
	\centering
	\includegraphics[width=0.49\textwidth]{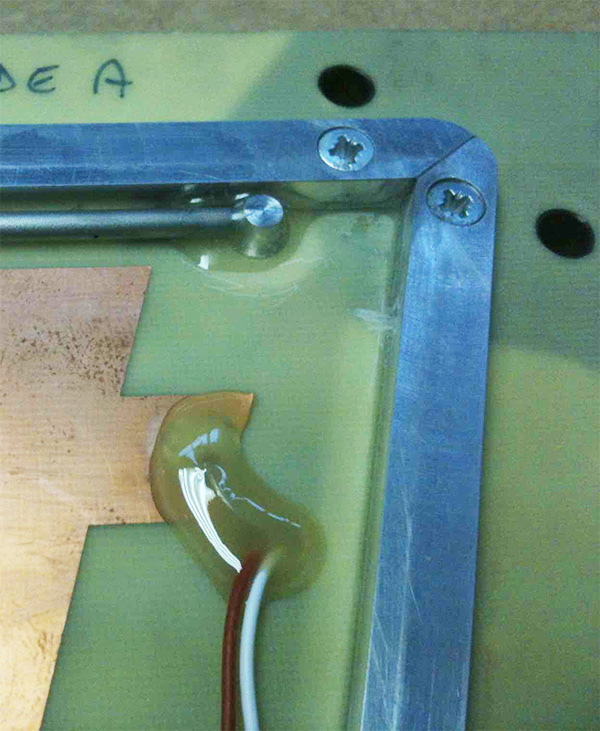}
	\includegraphics[width=0.485\textwidth]{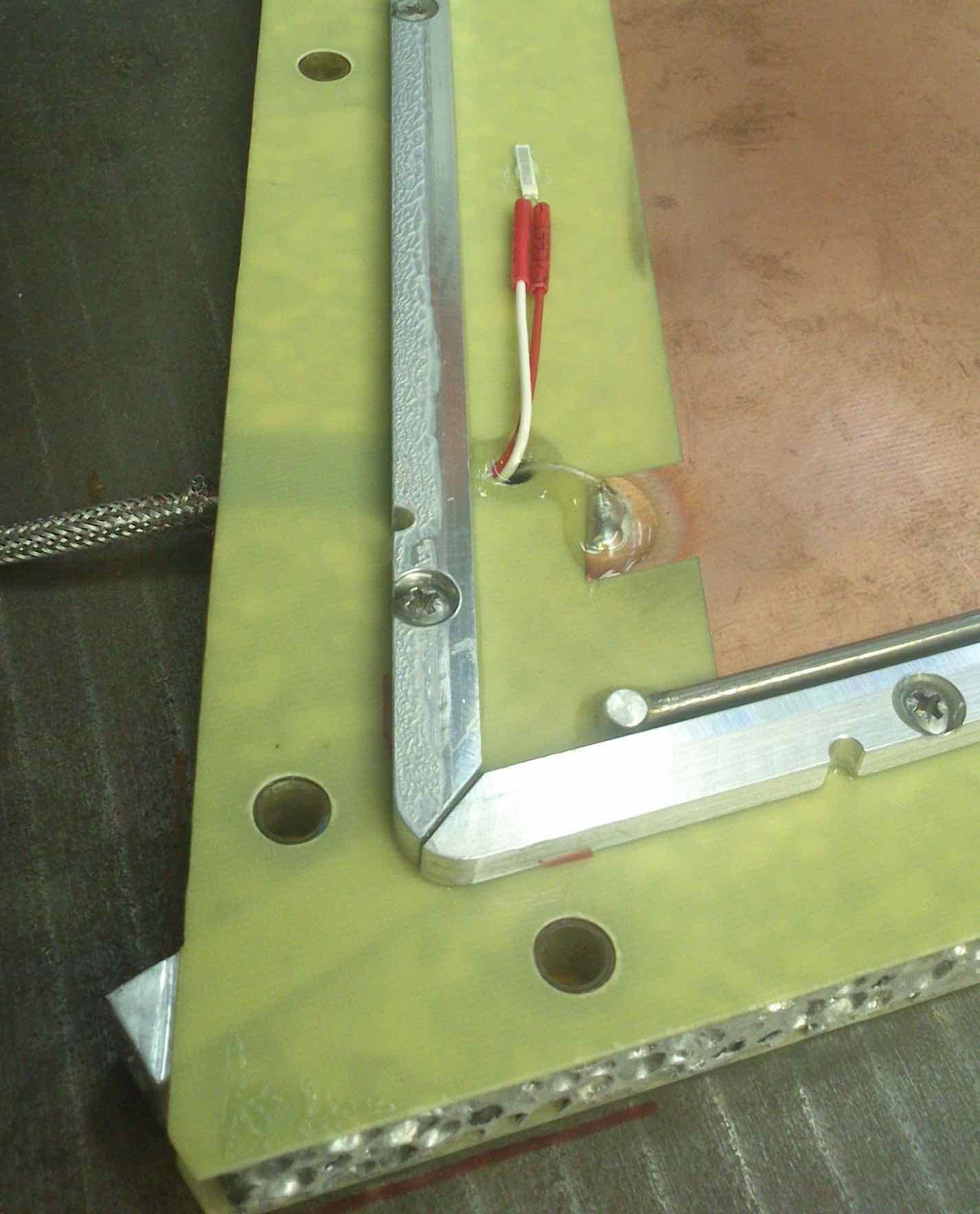}
	\caption{Photos showing details of the gas and HV feed-throughs of the drift panels as well as the cables for the temperature sensors. In the right photo also one of the openings in the mesh frame for the water evacuation is visible.}
	\label{Fig:DriftPanelServicesDetail}
\end{minipage}
\end{figure}

The drift electrode is connected to HV by an insulated wire, soldered to the Cu layer outside the active area. The wire is routed through a hole in the PCB into the panel and through another opening in the panel frame to the outside and sealed with Araldite, see Fig.~\ref{Fig:DriftPanelServicesDetail}.

In one of the two MMSW detectors PT1000 temperature sensors have been glued inside the drift volume but outside of the active area of the cathode (see detail in Fig.~\ref{Fig:DriftPanelServicesDetail}).

The resistive strips on the readout panels are connected to HV by soldering two thin insulated wires to the two HV contacts on each readout PCB\footnote{Note, that on the micromegas structures the resistive strips are split in the middle, creating two separate HV zones. The HV does not exceed 600~V.}. 

\subsubsection{Mesh glueing\label{sec:Mesh_glueing}}

The mesh material is woven stainless steel with 30~$\mu$m diameter wires and 50~$\mu$m openings (325 lines per inch). It is pre-stretched and glued on a transfer frame in industry. For a nominal tension of 10~N/cm, the measured mesh tensions show variations from 7 to 20~N/cm within the same mesh, acceptable for our needs. 

A thin layer of glue (Araldite 2011) is deposited on the chamfered surface of the mesh frame for the glueing of the mesh, leaving the flat surface of the mesh frame free from glue. A drop of glue is also deposited on the central insert.

\begin{figure}[htb]
    \begin{center}
        \includegraphics[width=0.36\textwidth]{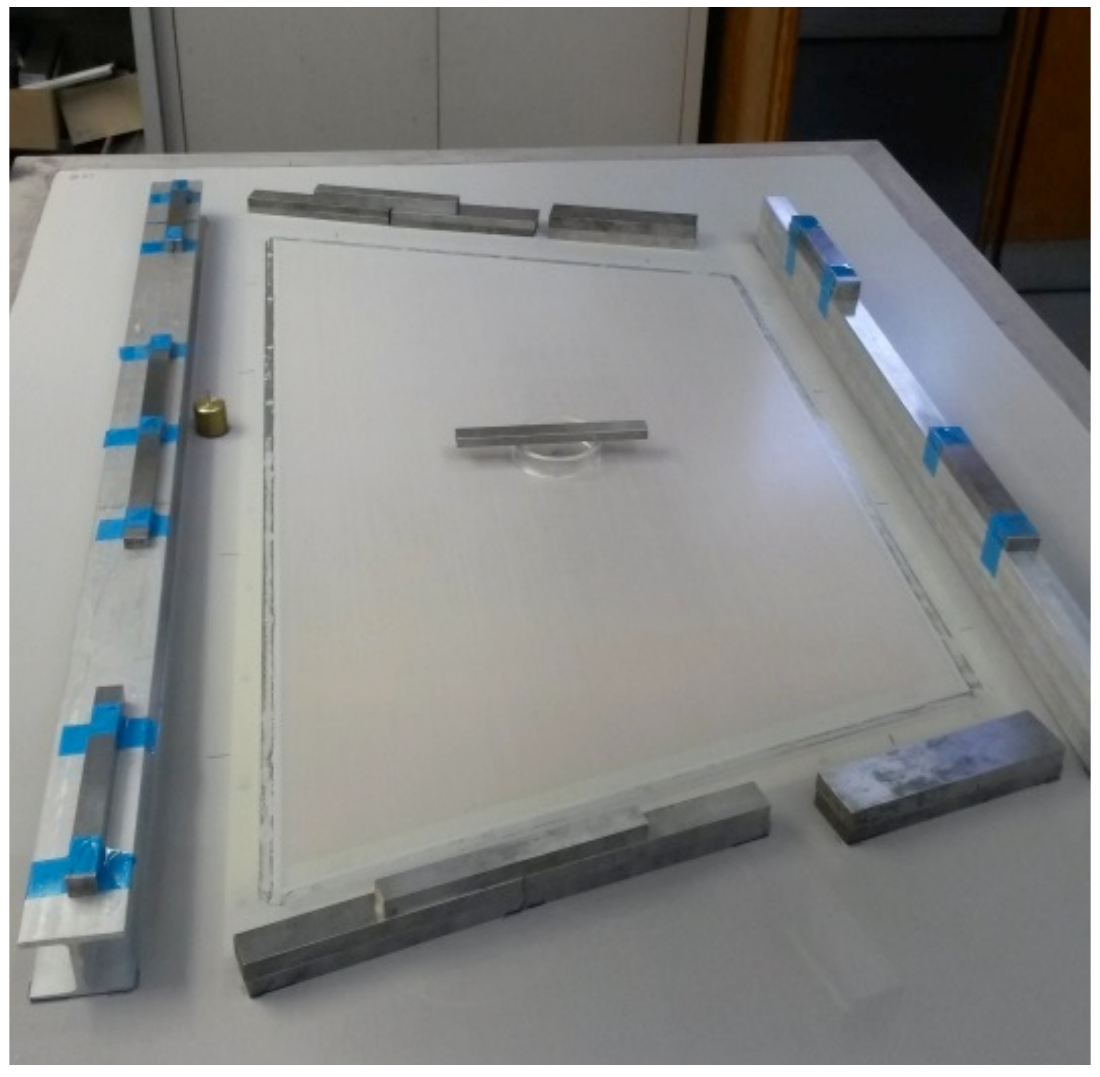}
        \includegraphics[width=0.62\textwidth]{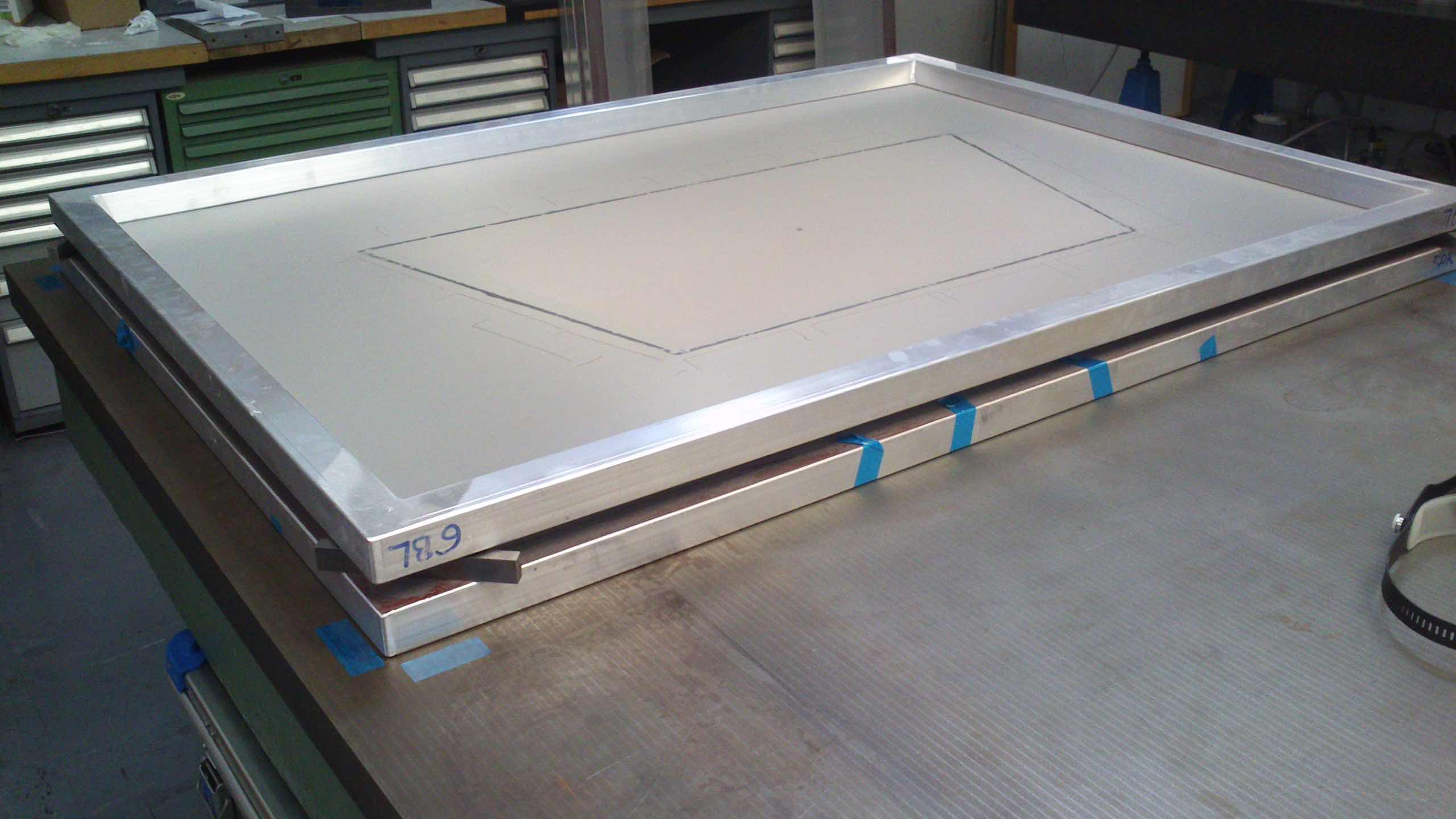}
        \caption{Photo showing the glueing of the mesh to its frame for the single mesh (left) and the double mesh case (right).}
        \label{Fig:MeshGluing}
    \end{center}
\end{figure}

The mesh, on its transfer frame, is then lowered onto the mesh frame in such a way that the perimeter of the mesh lays at a lower level with respect to the mesh frame. This configuration guarantees a good contact of the mesh at all the chamfered surfaces and a good adherence of the mesh.  Finally, some weights are placed on the outside of the mesh frame and around the central insert during the curing time of the glue (24 hours), shown in Fig.~\ref{Fig:MeshGluing}.

The mesh is only glued to the chamfered part of the frame. In this way, small glue excesses do not affect the stacking of the panels during the quadruplet assembly. Furthermore, a good electrical contact between the mesh and the 4~mm wide flat part of the frame is guaranteed.

After curing of the glue, the mesh is cut along the perimeter of the mesh frame and the transfer frame is removed. When the mesh is cut, its tension force is transferred to the mesh frame, resulting in panel deformations of O(0.5~mm)\footnote{The tightening screw through the central hole counter acts the deformation after assembly.}. Remeasuring the mesh tension, values very similar to the originals were found\footnote{Measuring the tension before the mesh was cut from the transfer frame, the tension was about 10\% higher than before. This is explained by the additional stretching of the mesh when it is placed on the frame. When the mesh was cut, these 10\% were lost again because of the relaxation until the mesh panel took its final form.}.

While the two outer drift panels have only one mesh, the central drift panel carries meshes on both sides. The glueing procedure follows the one of the single mesh. To prevent panel deformations after the first mesh has been glued the meshes were cut from their transfer frames only after the second mesh had been glued.  In this case the tension forces of the two meshes do approximately cancel and the panel stays flat.  

\subsubsection{Front-end electronics\label{sec:Front-end_electronics}}

Prior to the assembly of the quadruplets, the fixation of the front-end adaptor (mezzanine) boards is added to the panels.  

Zebra connector compression bars are screwed to the drift panel frames at the places where the front-end adaptor boards will be located, two bars on each outer drift panel and four bars on the central drift panel.

In addition, five ground connectors\footnote{Samtec MMCX-J-P-X-ST-SM1} for each adaptor board are soldered close to the outer edge of the readout board. They also serve as fixation for the mezzanine boards. 

\subsection{Mechanical quality control}

\subsubsection{Planarity}

\begin{figure}[htb]
    \begin{center}
        \includegraphics[width=0.49\textwidth]{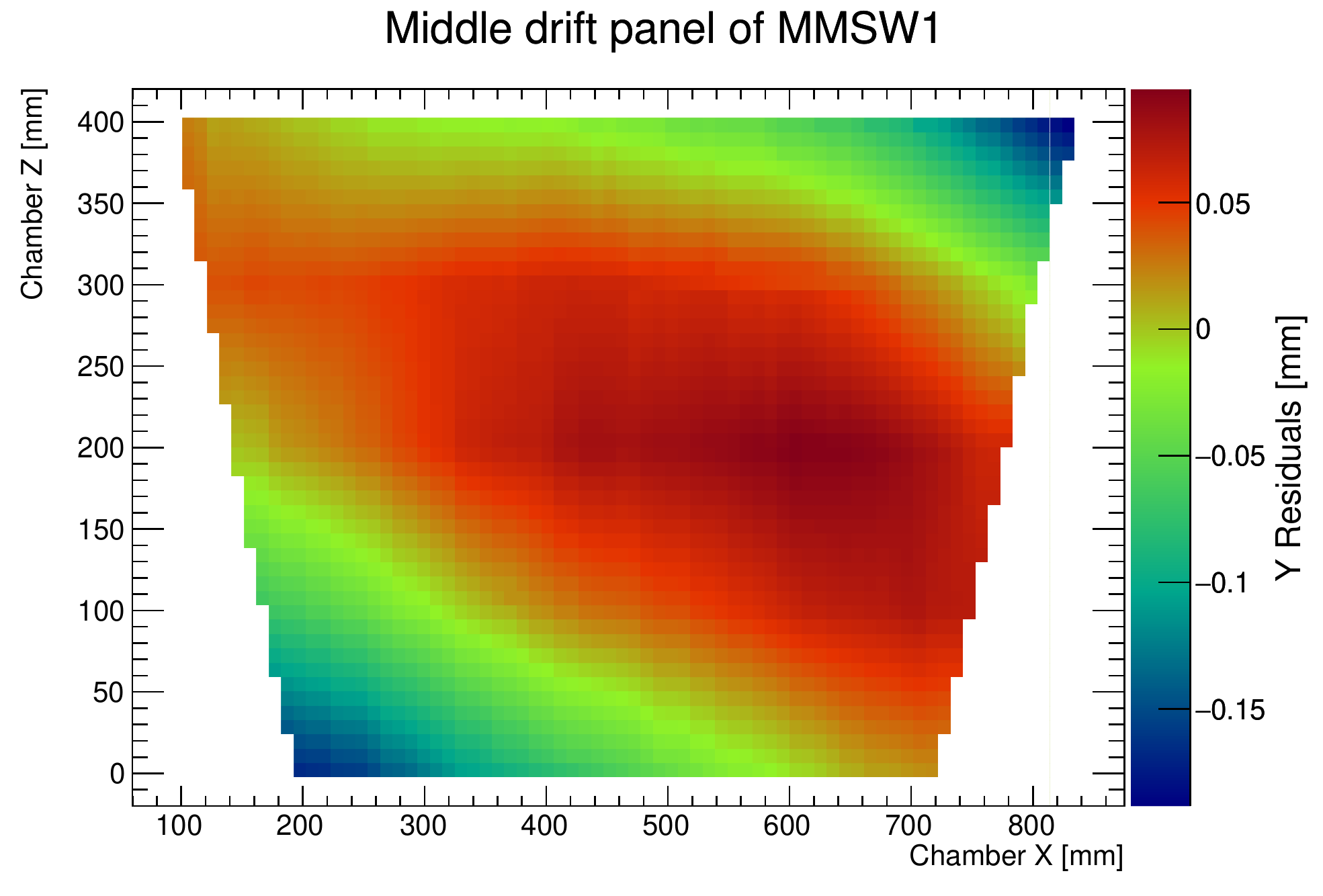}
        \includegraphics[width=0.49\textwidth]{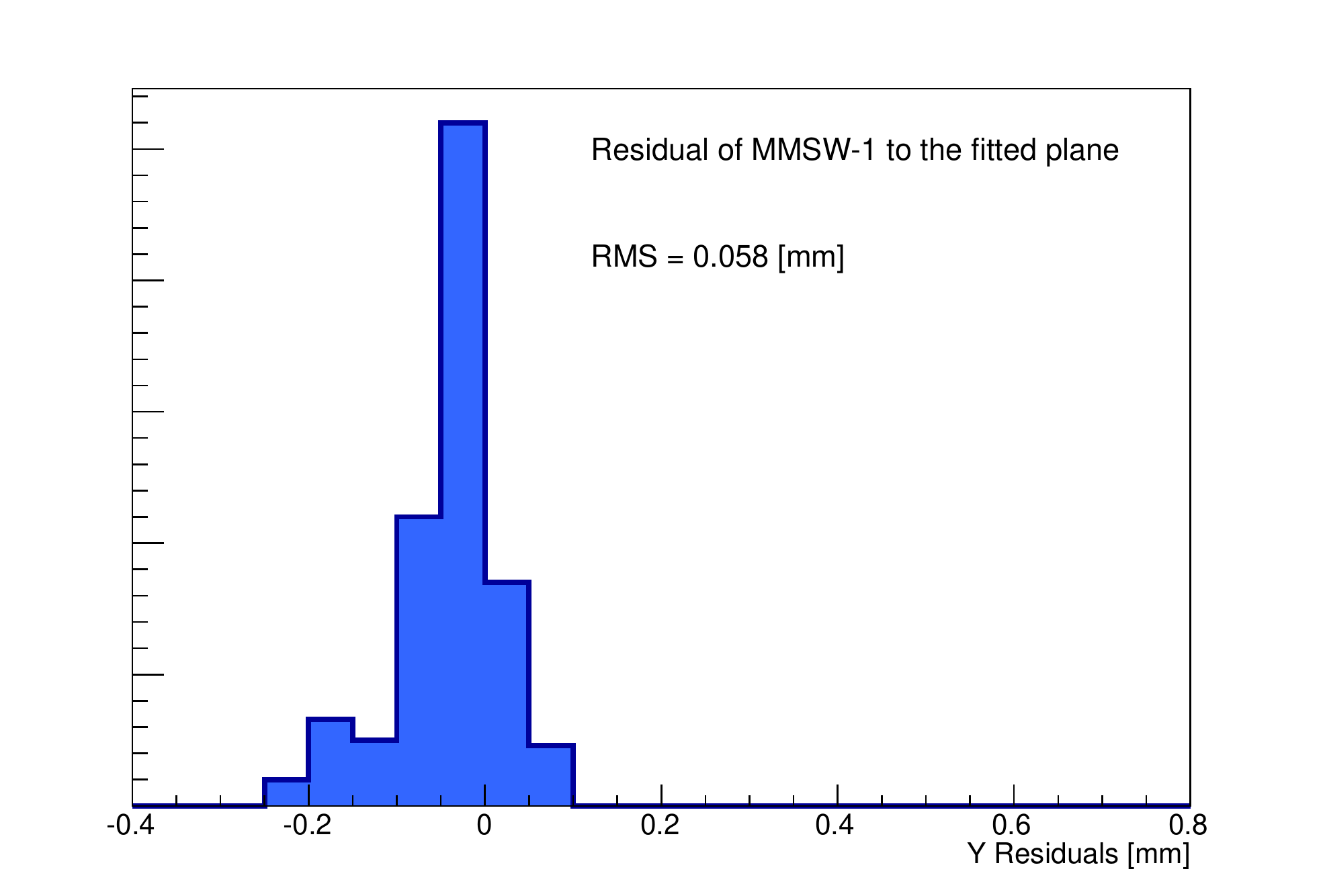}
        \caption{Left: Map of the drift panel surface before the mesh is glued, however, with the mesh frame mounted, right: distribution of the differences of each data point with respect to the fitted plane.}
        \label{Fig:DriftPlanarity}
    \end{center}
\end{figure}

The planarity of the panels was measured using a laser interferometer, resulting in a two-dimensional map of the panel surface. In order to subtract the relative misalignment of the panel with respect to the laser head, the data points were fitted with a plane. The displacements of the points from the fitted plane give the intrinsic shape of the panel, as shown in Fig.~\ref{Fig:DriftPlanarity} for the central drift panel of the first detector. 

The measurements show an overall flatness of 58$\mu$m (rms) with a bending in the bottom-left to top-right direction. The panel, as all the other drift panels, was  measured after the gluing of the mesh frame, which leads to additional stress, in particular for the external panels having the mesh frame on one side only.

\begin{figure}[htb]
    \begin{center}
        \includegraphics[width=0.49\textwidth]{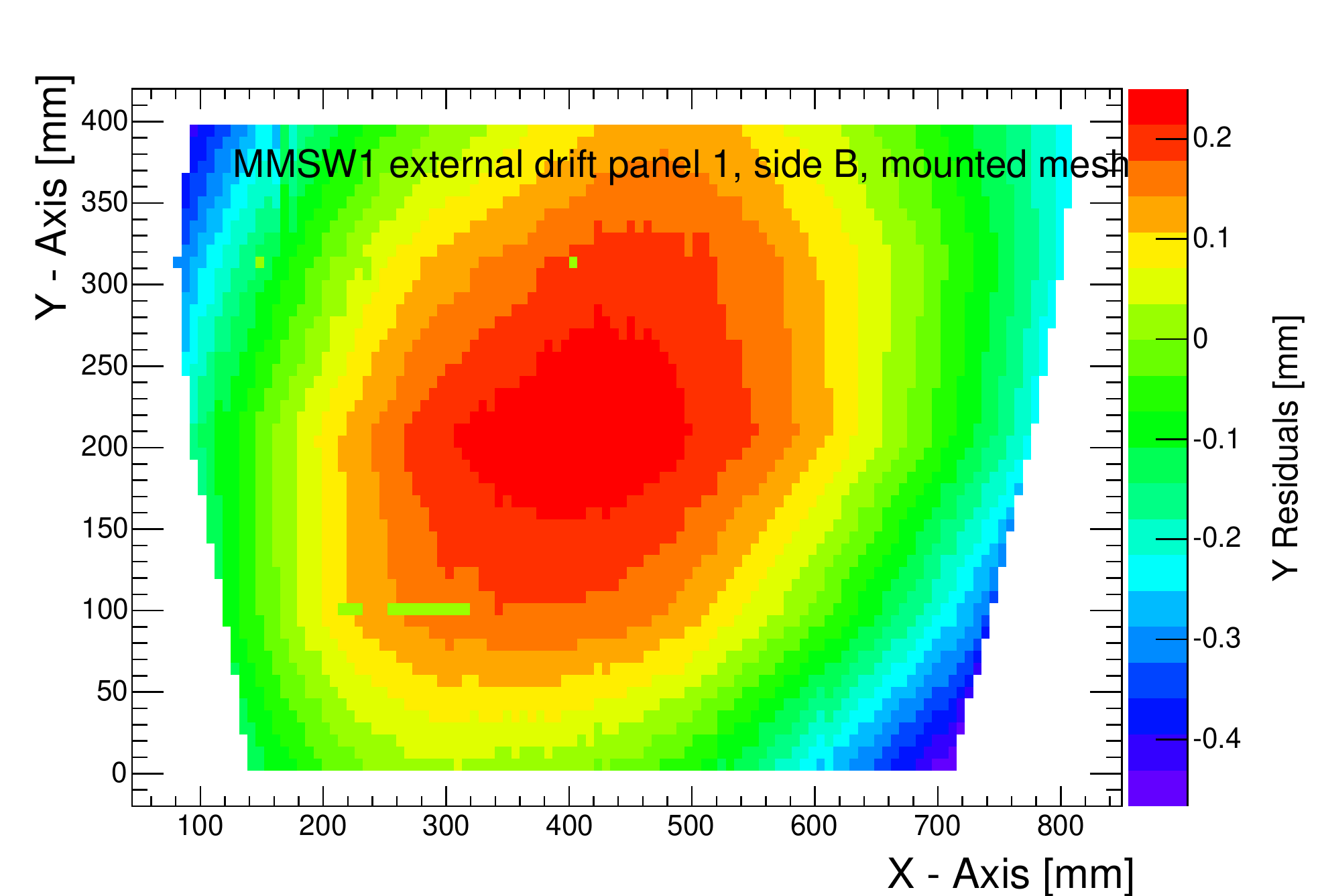}
        \includegraphics[width=0.49\textwidth]{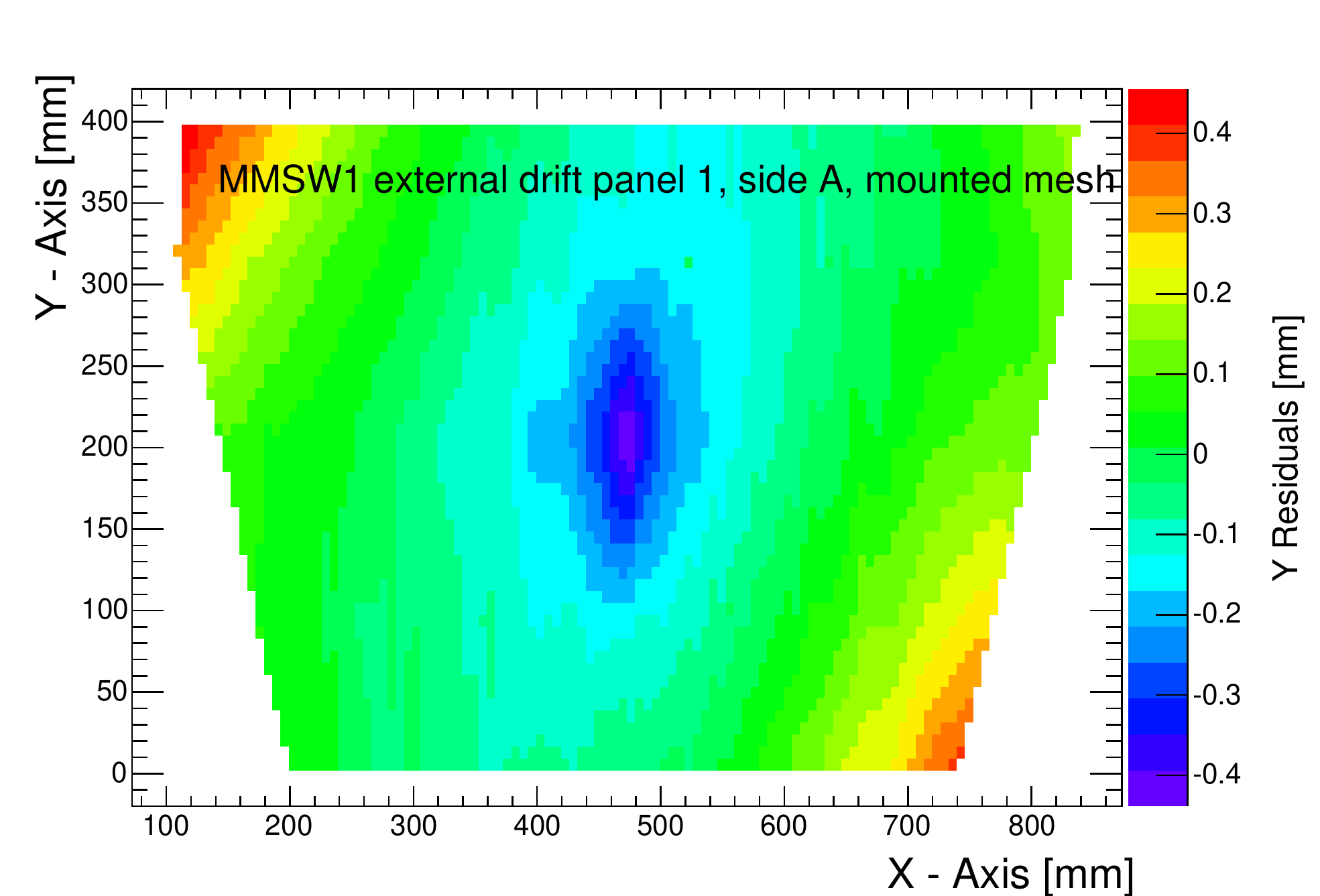}
        \caption{Map of the surfaces of an external drift panel equipped with the mesh. Left: the external surface, showing the panel bend. Right: the surface of the mesh, being constrained in the centre by the central spacer.}
        \label{Fig:DriftPlanarityWithMesh}
    \end{center}
\end{figure}

The drift panels were also measured after the gluing of the mesh, to quantify the bending introduced by the mesh. An example is shown in Fig.~\ref{Fig:DriftPlanarityWithMesh} (left) for one of the external panels. A bending up to 0.5 mm is measured in this case, with an RMS of 158$\mu$m. Figure~\ref{Fig:DriftPlanarityWithMesh} (right) shows the measurements obtained on the mesh surface for the same panel.  

Table \ref{Tab:PanelsPlanarity} summarises the rms values for the measured panels (all panels of MMSW-1 and readout panels only for MMSW-2). For the readout panels, the contribution to the rms owing to the complex structure of the mesh surface (described in  \ref{sec:PCB} and measured to be 33\,$\mu$m on average) is subtracted.

An average flatness of the symmetric panels (central drift and readout) of better than 60~$\mu$m is achieved. It should be noted that most of the measured panel deformations are going to disappear when the panels are constrained around the circumference and by the central connection after their assembly.

\begin{table} 
\begin{center}
  \begin{tabular}{ | c | c |c |}
    \hline
     Drift Panel &  Planarity (rms $\mu m$) &  Planarity (rms $\mu m$)   \\ 
         (MMSW1)             &  before mesh gluing         &  after mesh gluing   \\  \hline \hline
     Central Drift Side A & 58 & 59\\ 
     Central Drift Side B & 39 & 53\\     
     Ext Drift 1 Side A & 136 & 113 \\ 
     Ext Drift 1 Side B & 103 & 158\\ 
     Ext Drift 2 Side A & 81 & 105\\ 
     Ext Drift 2 Side B & 104 & 138\\ \hline \hline
     Readout Panel &  Flatness (rms $\mu m$) &  Flatness (rms $\mu m$) \\ 
                            &  MMSW1                           & MMSW2 \\ \hline \hline
     Readout Eta Side A & 39 & 63 \\  
     Readout Eta Side B & 42 & 64 \\  
     Readout Stereo Side A & 83 & 62 \\  
     Readout Stereo Side B & 75 & 70 \\  
    \hline
  \end{tabular}
   \captionof{table}{Summary of RMS of the measured panel flatness after different construction steps.}
     \label{Tab:PanelsPlanarity}  
   \end{center}
   \end{table}

\subsubsection{Alignment}

\begin{figure}[t]
    \begin{center}
        \includegraphics[width=0.52\textwidth]{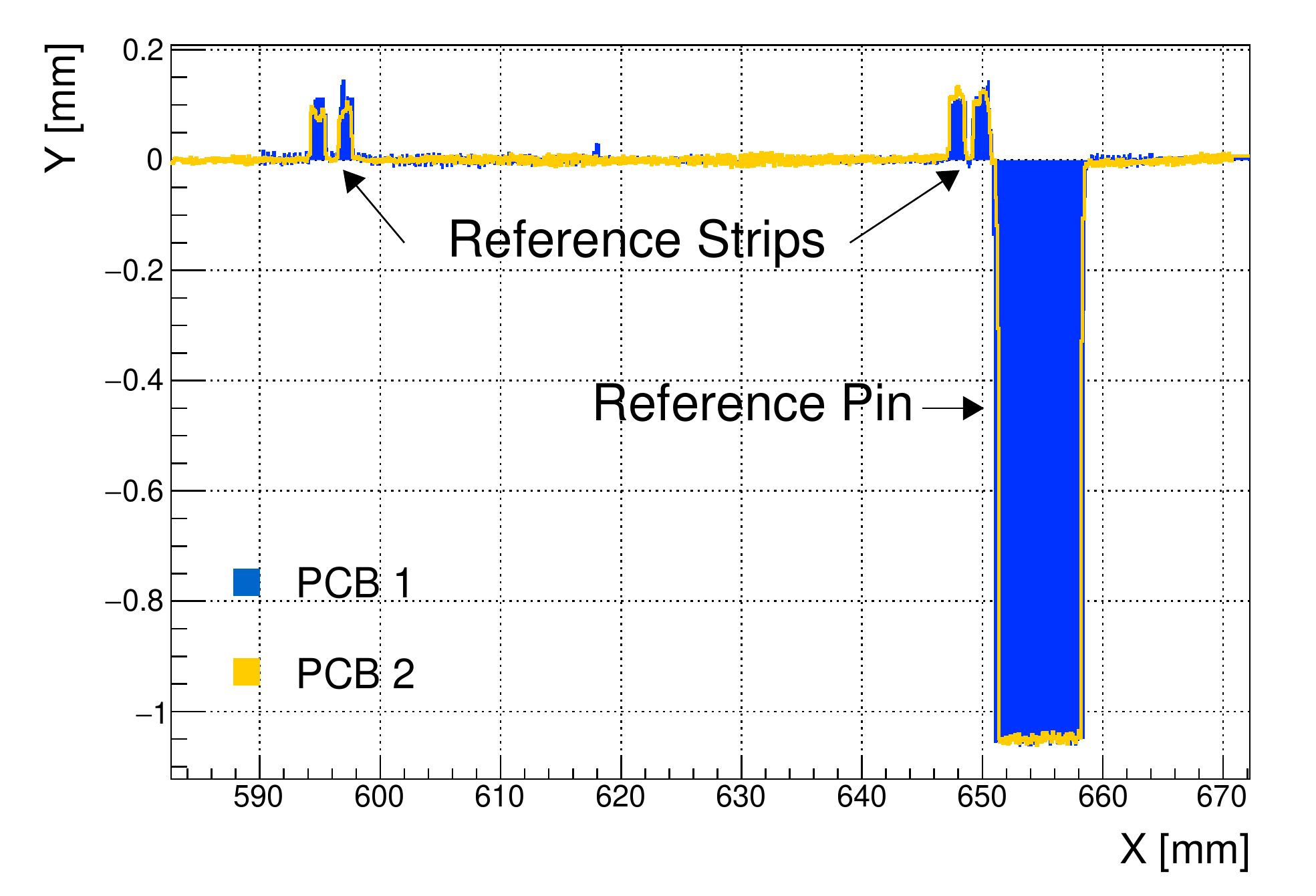}
        \includegraphics[width=0.47\textwidth]{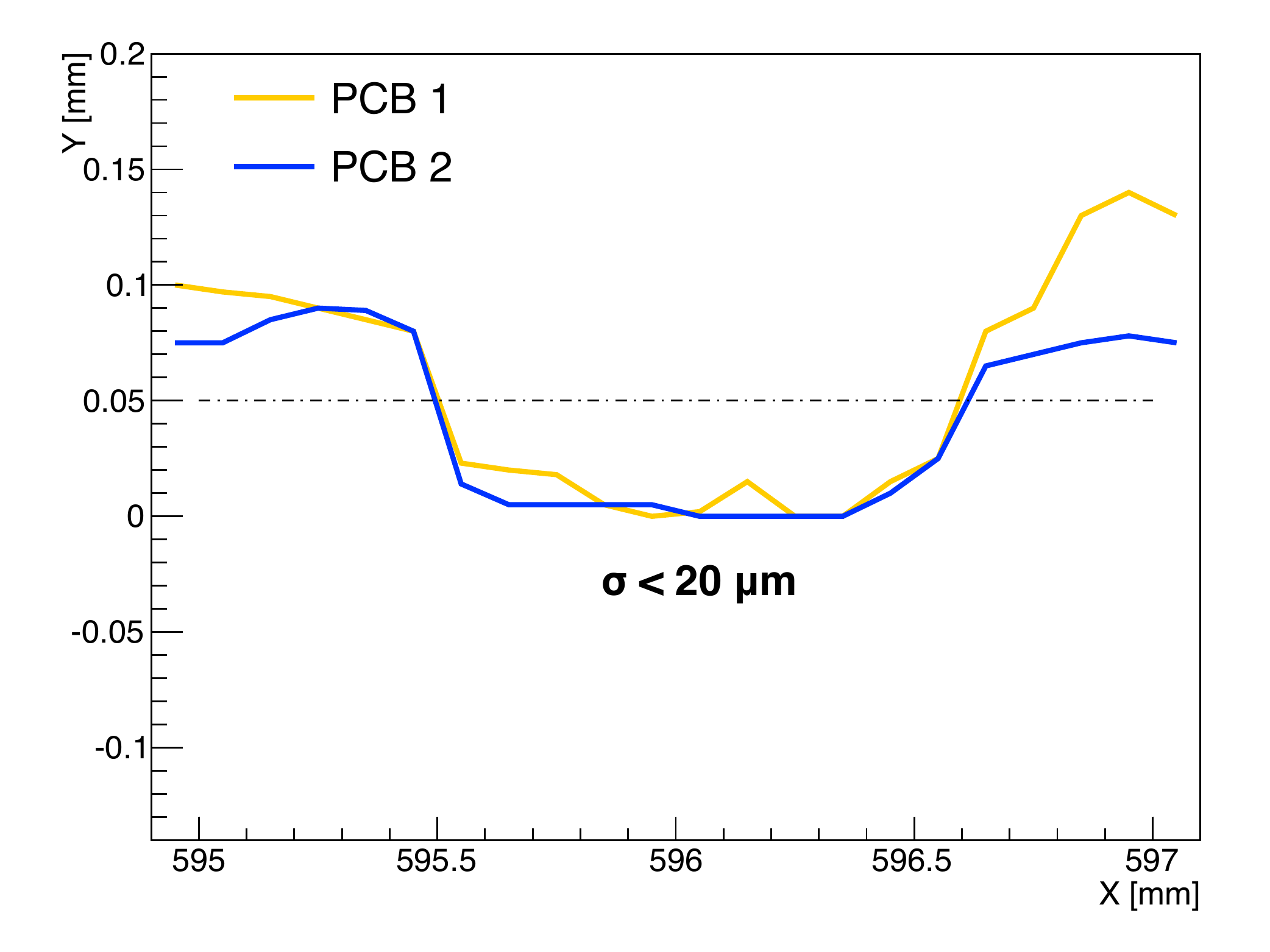}
        \caption{Left: Measurements acquired with the laser tracker. Blue and orange data points refer to the top (PCB1) and bottom (PCB2) sides of one panel of the first prototype, respectively. The centres of the precision pin in the two measurements have been aligned. Right: Zoom of the measurements in the region of two alignment strips. For the relative alignment of the strip patterns an upper limit of $20\,\mu \mathrm{m}$ can be set.}
        \label{Fig:StripAlignment2}
    \end{center}
\end{figure}

While the alignment of the two panel faces is not critical for the drift panels, it is of prime importance for the readout panels. The goal is to know the strip positions to better than 40~$\mu$m on all detection layers. To measure the relative strip positions on the two faces of the readout panels we take advantage of the readout strips that are routed to the side of the readout panel (two out of each group of 128), see Fig.~\ref{Fig:ReadoutBoards}. Inserting a precise pin with a 10~mm diameter head in one of the close-by peek inserts the distance between pin and strips has been measured, again using the laser interferometer. The same procedure was repeated for the opposite PCB where the reference pin was inserted from the opposite side. The data were analyzed by aligning the centre positions of the reference pin on the two sides and checking the alignment of the strips. The result is shown in Fig.~\ref{Fig:StripAlignment2}. For both detectors the readout strips on the two sides of the $\eta$ panels are aligned to better than 20~$\mu$m.

\subsection{Quadruplet assembly\label{sec:Quadruplet_assembly}}

\begin{figure}[t]
    \begin{center}
        \includegraphics[width=0.99\textwidth]{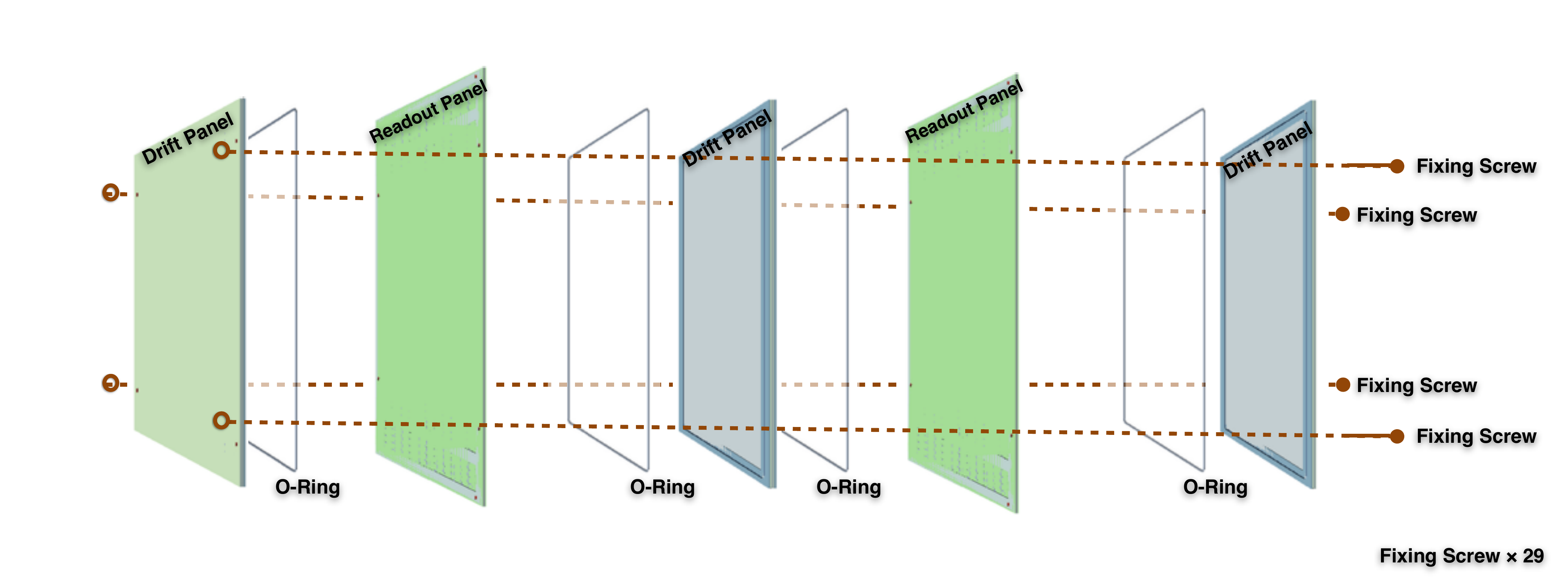}
        \caption{Schematic illustration of the panel assembly procedure of the two MMSW prototypes.}
        \label{Fig:DetectorAssembly}
    \end{center}
\end{figure}

The assembly sequence of the quadruplet is illustrated in Fig.~\ref{Fig:DetectorAssembly}. The gas tightness is ensured by four O-rings\footnote{EPDM DT-10 2621 0506}, that are inserted between the drift and readout panels around the mesh frames. The distance between the panels is defined by gas gap spacers, consisting of four 5~mm thick precision-machined bars with assembly holes. 

In order to avoid dust particles entering the drift or amplification region, the assembly of the quadruplet is performed in a clean environment. Dust particles are removed from the readout panels and the meshes using a clean room vacuum cleaner, soft flow of clean compressed air, and a rubber roller passed on sticky paper. Next, the first drift panel is placed on the table, with the mesh facing upwards. Four of the assembly screws are inserted as guides in the corners and the gas gap spacer bars are mounted. Then the first readout panel is added from the top, for now without the O-ring. Weights are added around the perimeter of the readout panel to guarantee a good contact. The electrical integrity of the assembly is tested by measuring the insulation between the resistive strips and the mesh applying 500~V to the strips. Any short is immediately visible. In this case the panel assembly is opened and re-cleaned. The assembly is declared as electrically sound when resistance values of several tens of G$\Omega$ are reached after sufficient charge-up time\footnote{Because of the large capacitance of the assembly it may take a minute to reach these values.}. 

This procedure is repeated for the other panels. At the end the O-rings are mounted, gap by gap. For this, the corresponding gas gap spacer bars are removed. At this point the mesh frame (that is slightly thinner than the gas gap frame) defines the distance between the two panels and "seals" the inside. The O-ring is pushed between the two panels, removing and replacing the four assembly screws in the corners one by one. Then the gas gap bars are replaced. 

Before tightening the assembly bolts two precision alignment pins are inserted into the precision peek inserts. In a final step, the assembly is tested again for electrical integrity.

\subsection{Quadruplet services\label{sec:Quadruplet services}}

The quadruplet is equipped with two rows of SHV HV connectors to which the HV wires of the four readout panels are connected. Another SHV connector distributes the HV to the four drift electrodes (Fig.  \ref{Fig:FinalDetector}). Four manifolds for the gas input and output lines are added providing for parallel gas flow through the four gas gaps. After that, the mezzanine boards with their front-end electronics are mounted.

\subsection{Design and construction improvements\label{sec:Improvements}}

During the construction of MMSW-1 \cite{Iengo:2014vxa}, several difficulties in the construction had been encountered, leading to an improved design for the construction of MMSW-2. The change in the vacuum table construction has been already discussed in Sect.~\ref{sec:VacuumTable}.

The other major improvement was the simplification and re-design of the gas gap spacer, gas distribution, and mesh frame. In the first design all functions were combined in the gas gap spacer. It consisted of a single 5~mm thick profile that served as mesh frame but also integrated the gas input and output channels, the gas manifolds, and an O-ring groove (the O-ring was only 2~mm in diameter). What looked attractive at first sight turned out to be much more complicated and time-consuming in machining and assembly.  

In the re-design, the gas supply system of the MMSW-2 has been completely reviewed, leading to the design described above\footnote{This design has in the meanwhile be adapted as baseline for the NSW micromegas detectors.}. 

After the construction of and first experience with MMSW-2, MMSW-1 has been refurbished to follow the new design. In particular, the two external drift panels were made by replacing the aluminium honeycomb and the panel frames with aluminium foam of density 0.3~g/cm$^3$. The resulting panels have  a similar planarity as the honeycomb-based panels and similar stiffness, however, are about 10\% lighter. The advantages and disadvantages in the panel construction are about neutral. The time and costs that were saved in not needing the panel frames were made up by the need to machine the Al foam panels to achieve the reported panel flatness. In addition, solid inserts had to be added to the foam panels where the Zebra connector compression bars needed to be fixed to the drift panel.

\section{\label{sec:Performance}Basic Performance Studies}

The MMSW detectors have been tested in test-beam, X-Ray, and cosmic ray measurements. We will concentrate in the following on the main performance parameters, such as efficiency, gain homogeneity, and spatial resolution. 

The MMSW detectors have been read out through the RD51 Scalable Readout System (SRS)~\cite{Martoiu:2013aca,Martoiu:2011zja}. The final electronics for the NSW micromegas not yet being available, APV25 (Analog Pipeline Voltage chips with $0.25\,\mu\mathrm{m}$ CMOS technology) \cite{APVJones} hybrid cards are used instead. The APV hybrid cards are mounted on intermediate (mezzanine) boards that are connected to the readout panels through Zebra connectors\footnote{The APV hybrid cards carry 130-pin connectors, not compatible with the Zebra connection scheme chosen for the MMSW.}. Each mezzanine hosts four APV25 boards and covers 512 strips.

\begin{figure}[h]
    \begin{center}
        \includegraphics[width=0.5\textwidth]{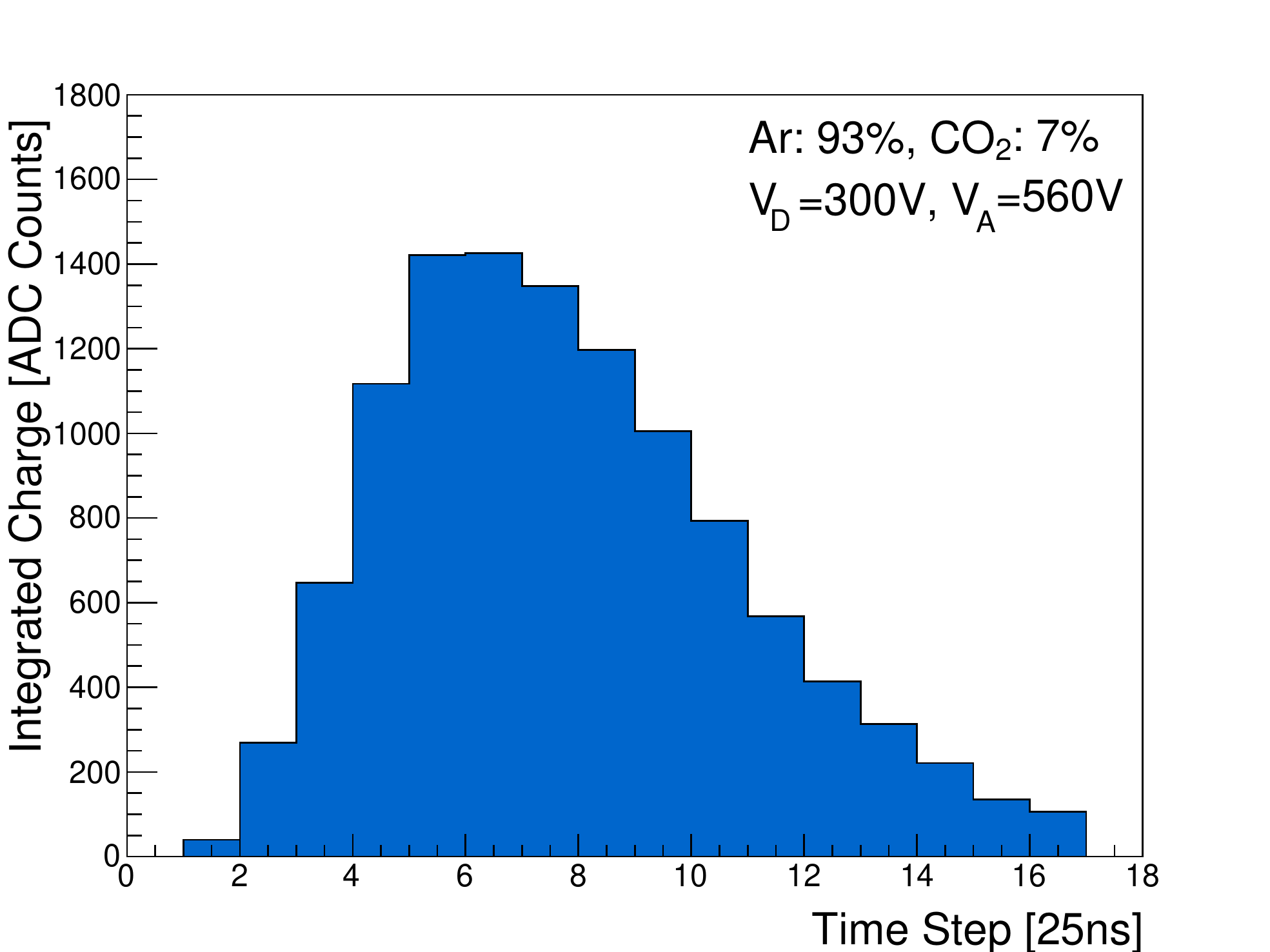}
        \caption{Example of an APV25 output signal on one readout strip, showing the development of the integrated charge for 18 time bins of 25~ns.}
        \label{Fig:APVsignal}
    \end{center}
\end{figure}

The APV25 chip has 128 channels and delivers analog CR-RC shaped signals sampled at 40 MHz. The analog signals from the APV25 are transmitted via HDMI cables to the SRS ADC card where they are digitised and transmitted to the data acquisition PC. Here, a zero suppression algorithm is applied that removes 'empty' data.  Since the APV25 chip had been developed for the readout of silicon detectors it uses a short integration constant that is not optimised for the micromegas with a longer charge collection time. Another drawback is the limited dynamic range of the chip. We use the chip in a mode that samples the integrated charge over up to 27 time bins of 25~ns. An example of the APV25 output for a single electronics channel, after zero suppression is given in Fig.~\ref{Fig:APVsignal}. It shows the development of the integrated charge in 25~ns bins. From these data, for each channel, the signal charge and a time stamp are extracted. The signal charge is defined as the maximum of the integrated charge samples. The time stamp is taken as the maximum of the first derivative of the rising edge of the signal (or the half rising time). 

\subsection{Event reconstruction and efficiency measurements}

The performance studies based on cosmic muons were performed in the ATLAS cosmic ray stand in the GDD (Gaseous Detector Development) laboratory at CERN \cite{Chefdeville:2011zz}. Two layers of twelve plastic scintillator slabs with an active area of $2.5\times1.1$\,$\mathrm{m}^2$ in each plane at a distance of 2\,m have been used as coincidence trigger. The MMSW has been operated with a gas mixture of Ar:CO$_2$ (93:7) at atmospheric  pressure, amplification voltages ranging from $V_\mathrm{A}=500\,\mathrm{V}$ to $V_\mathrm{A}=600\,\mathrm{V}$, and a fixed drift voltage of $V_\mathrm{D}= -300\,\mathrm{V}$. 

\begin{figure}[t]
    \begin{center}
        \includegraphics[width=0.49\textwidth]{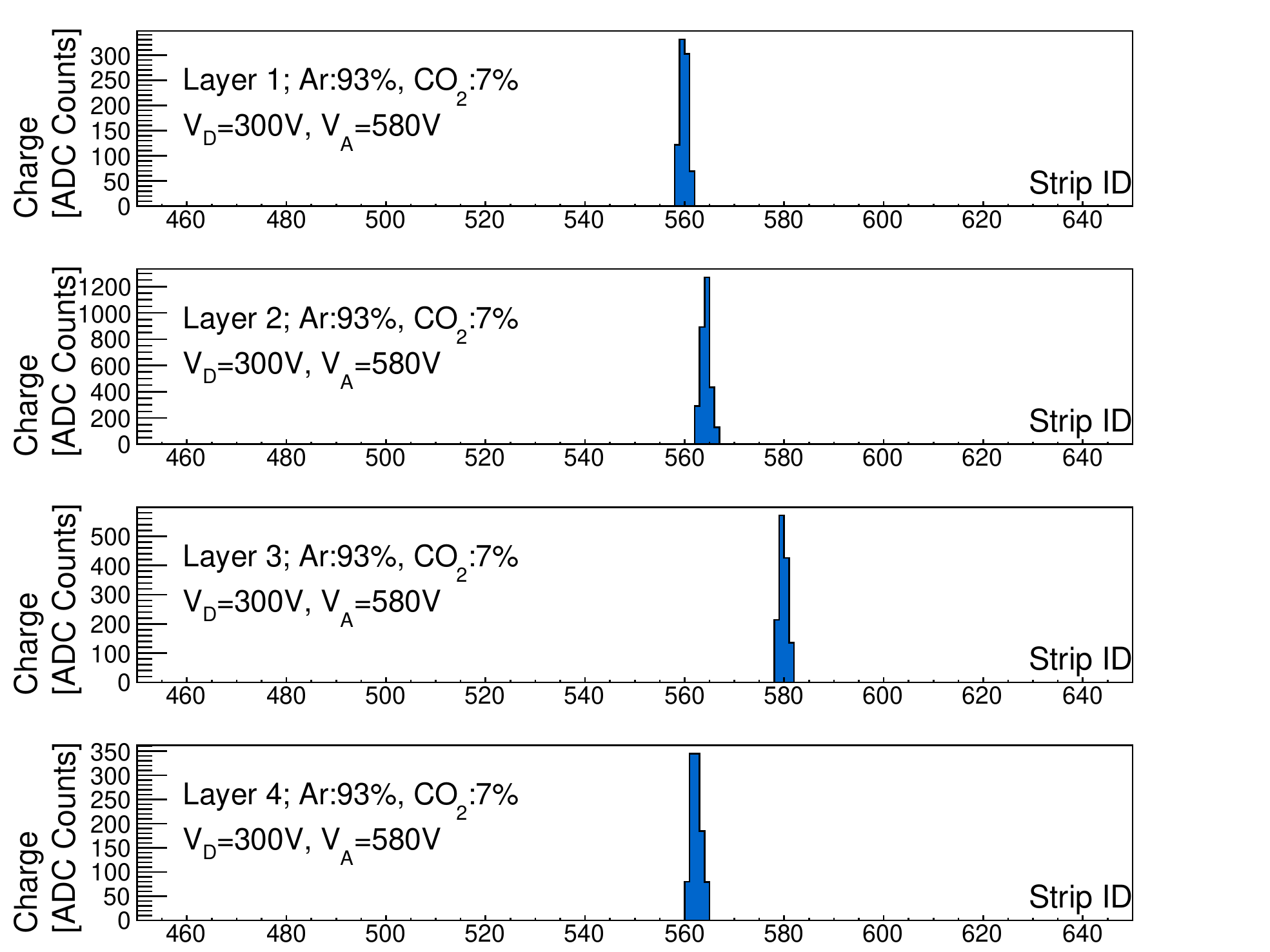}
        \includegraphics[width=0.49\textwidth]{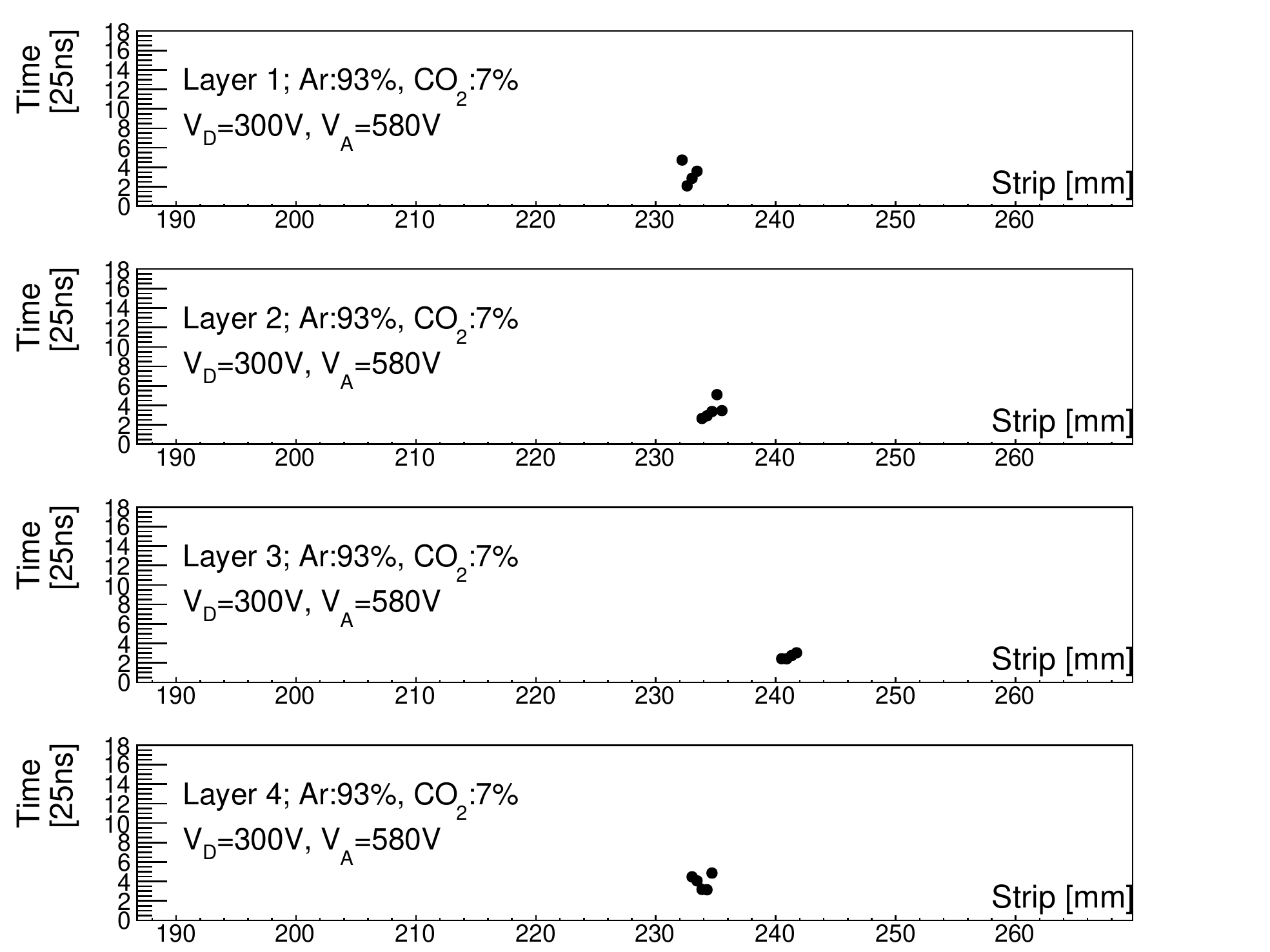}
        \caption{Single cosmic muon event recorded in four layers of MMSW. The left plots shown the strips maximum charge vs. strip ID among all time bins.  The right plot shows the signal arrival time vs. strip position}
        \label{Fig:SingleTrackingEvent}
    \end{center}
\end{figure}

Figure \ref{Fig:SingleTrackingEvent} shows a typical cosmic muon event where the charge and time for a single particle passing through the four layers of the MMSW is displayed. Clean hits are observed in each layer, characterized by a cluster of neighbouring strips. Owing to the stereo angles of the strips in the third and forth readout layer the hit positions (in term of strip number) in these layers are shifted with respect to the ones in the $\eta$ layers. The more the track position is distant from the centre line of the detector the more the hits are shifted. 
 
A cluster of strips is defined by the following criteria:

\begin{itemize}
\item it must contain at least three readout strips
\item the strips are contiguous and not separated by more than one missing strip. This requirement is important to allow for faulty readout strips or problems in the readout electronics
\item the sum of all strip charges (cluster charge) must be above a minimal threshold
\end{itemize}

The cluster position is defined as the charge-weighted average position of all strips of the cluster. This definition, strictly speaking, holds only true for perpendicular incoming particles. However, the corresponding systematic uncertainties for incoming particles under different angles can be neglected for the following studies.

\begin{figure}[t]
    \begin{center}
	\includegraphics[width=0.49\textwidth]{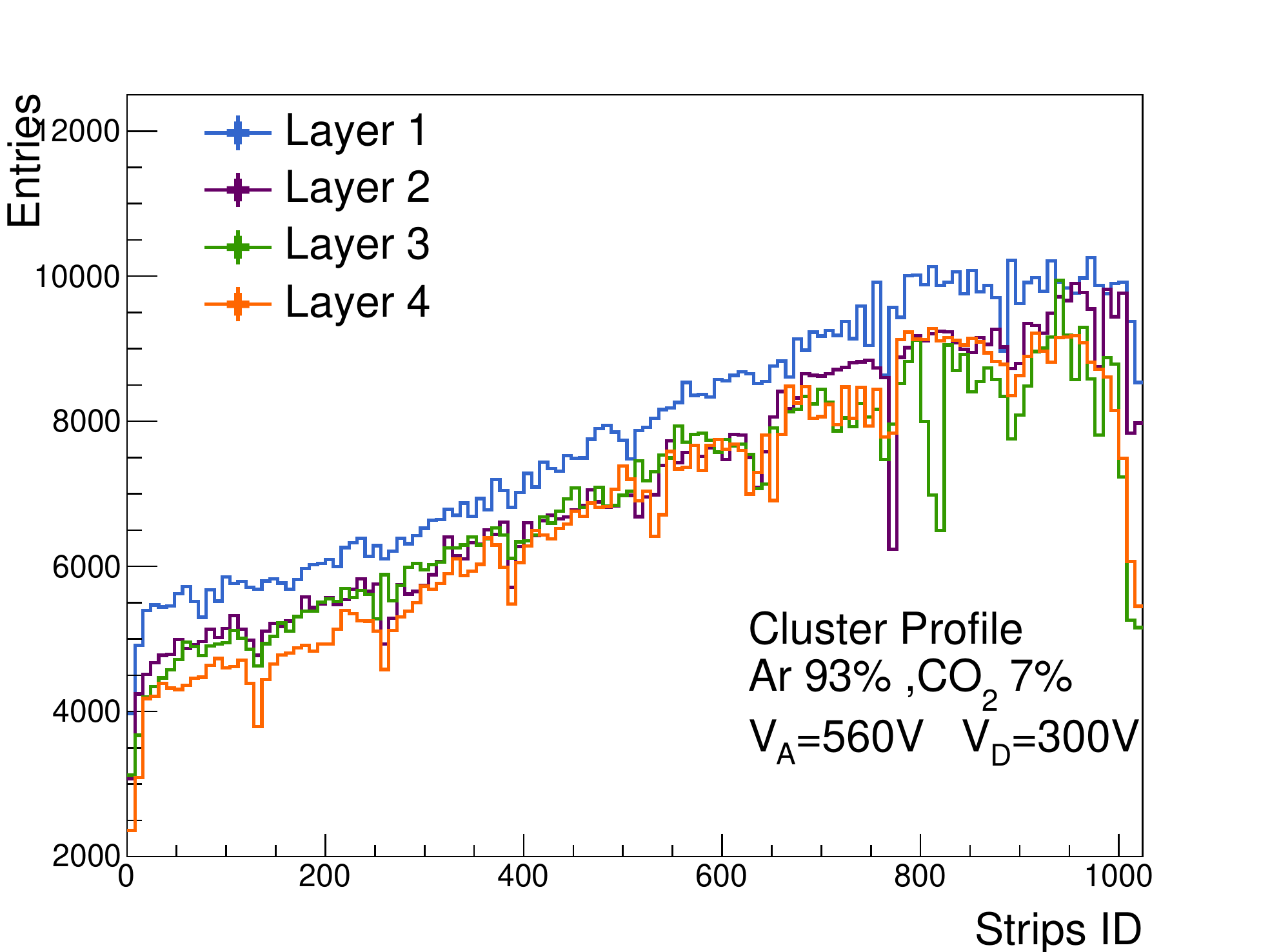}
	\includegraphics[width=0.49\textwidth]{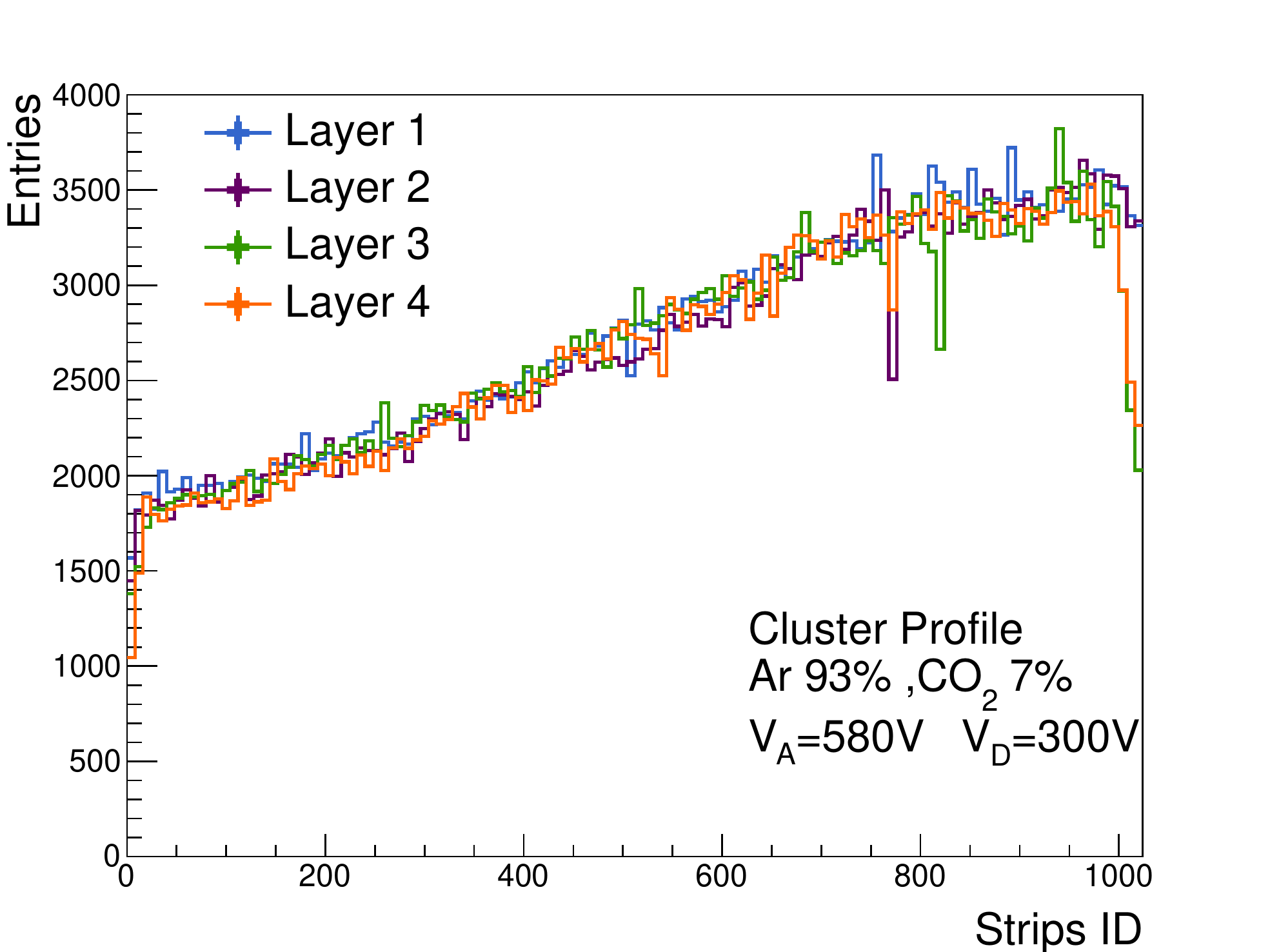}
	\caption{Number of clusters as function of the strip number for all layers for an amplification voltage of $V_A$=560\,V (left) and $V_A$=580\,V (right).}
	\label{Fig:HitMap}
    \end{center}
\end{figure}

The number of clusters as a function of the readout strip position for the four detector layers is shown in Fig. \ref{Fig:HitMap} for $V_\mathrm{A}=560$\,V and $V_\mathrm{A}=580$\,V with $V_\mathrm{D}=-300$\,V, respectively. A similar count rate is observed for all layers at $V_\mathrm{A}=580$\,V, while the full efficiency for the layers 2-4 is not yet reached at $V_\mathrm{A}=560$\,V.

The increase of the number of clusters with increasing strip number in all four layers is a geometrical effect. About half of the effect comes from the length increase of the readout strips, the other half can be attributed to the efficiency of the scintillators used for triggering. 
The decrease at very high and very low strip numbers in the third and fourth layer is the result of the shorter strips at the detector edges. 

\begin{figure}[htb]
    \begin{center}
        \includegraphics[width=0.49\textwidth]{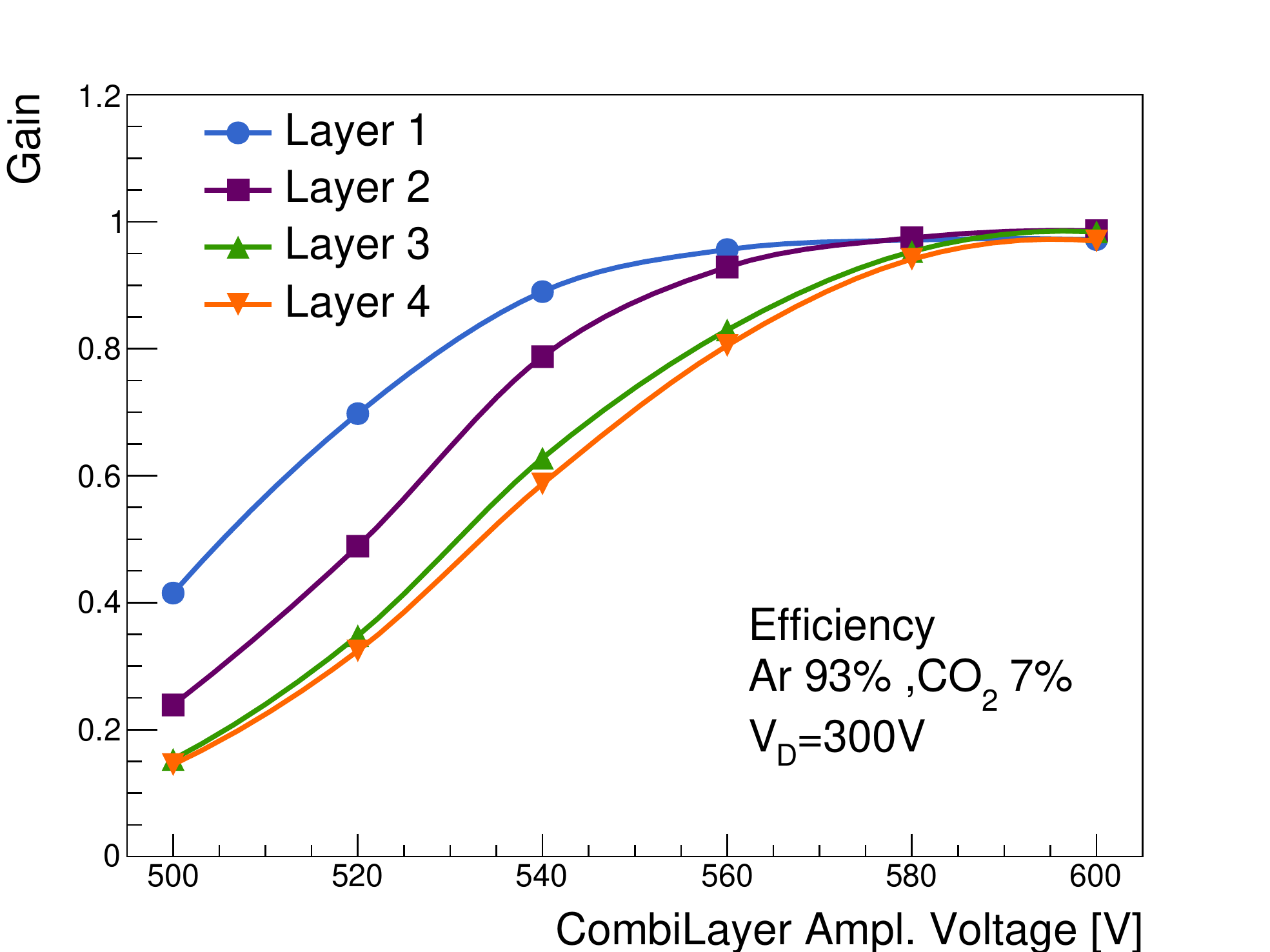}
        \includegraphics[width=0.49\textwidth]{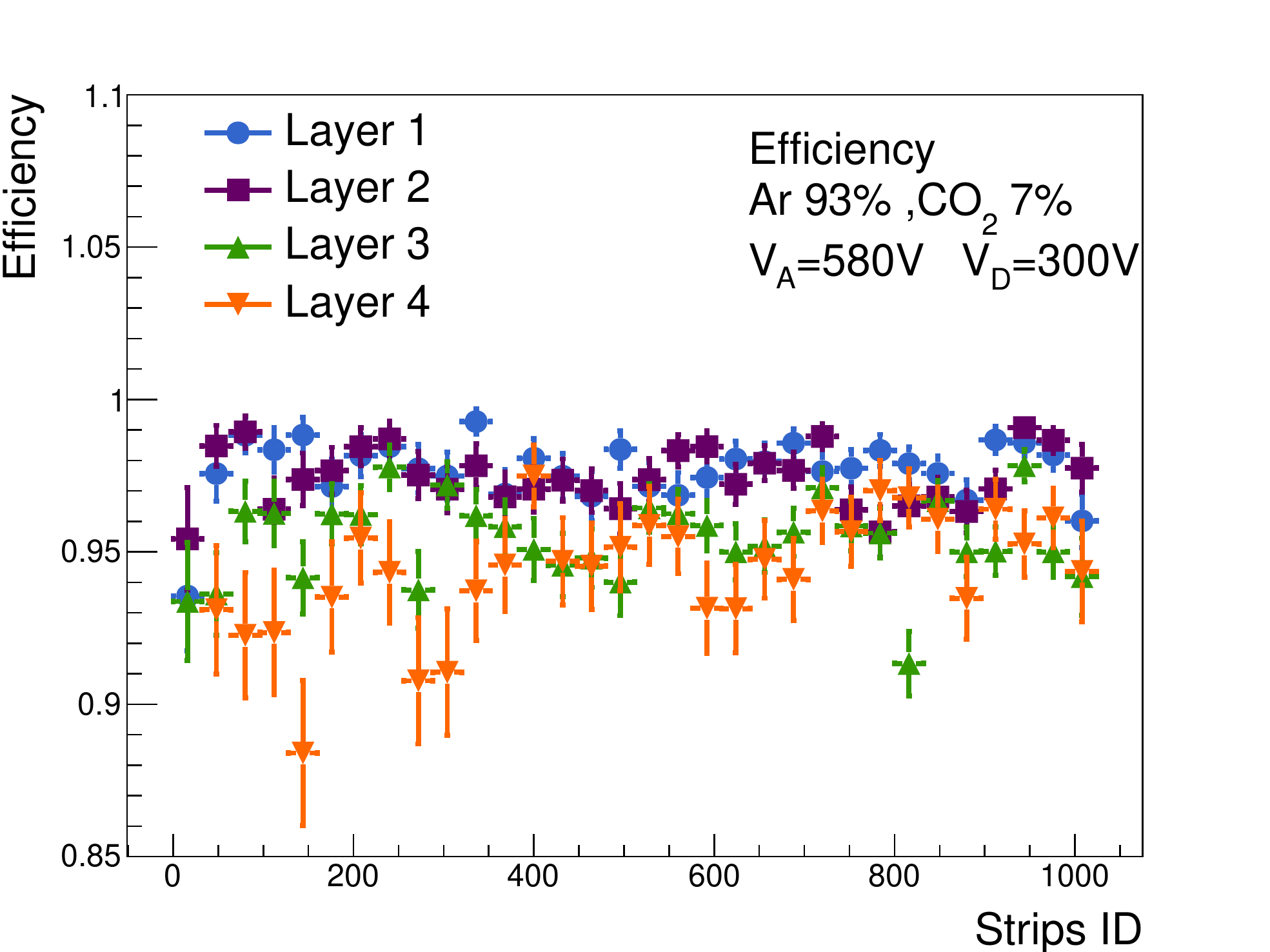}
        \caption{Layer efficiency vs. the amplification voltage (left) and cluster reconstruction efficiency vs. strip position for $V_\mathrm{A}=580\,$V for all layers (right).}
        \label{Fig:Efficiency}
    \end{center}
\end{figure}

To study the efficiency of the MMSW, we define three out of four readout layers as reference layers and test the remaining layer. We require exactly one cluster per reference layer and apply a tracking algorithm for the prediction of the cluster position in the test layer. A cluster is defined as correctly reconstructed if its position is within 2.5\,mm of the predicted position.
The dependence of the detector efficiency on the amplification voltage and the efficiency vs. the  strip ID for $V_\mathrm{A}=580\,$V is shown in Fig. \ref{Fig:Efficiency}. Full efficiency is reached at 560~V for the first layer, at about 570~V for second layer, and at 580~V for the third and fourth layers.
For amplification voltages higher than $V_\mathrm{A}=580$\,V, the reconstruction efficiency is above 95\% for all layers and uniform across all readout strips.  

\begin{figure}[htb]
    \begin{center}
        \includegraphics[width=0.49\textwidth]{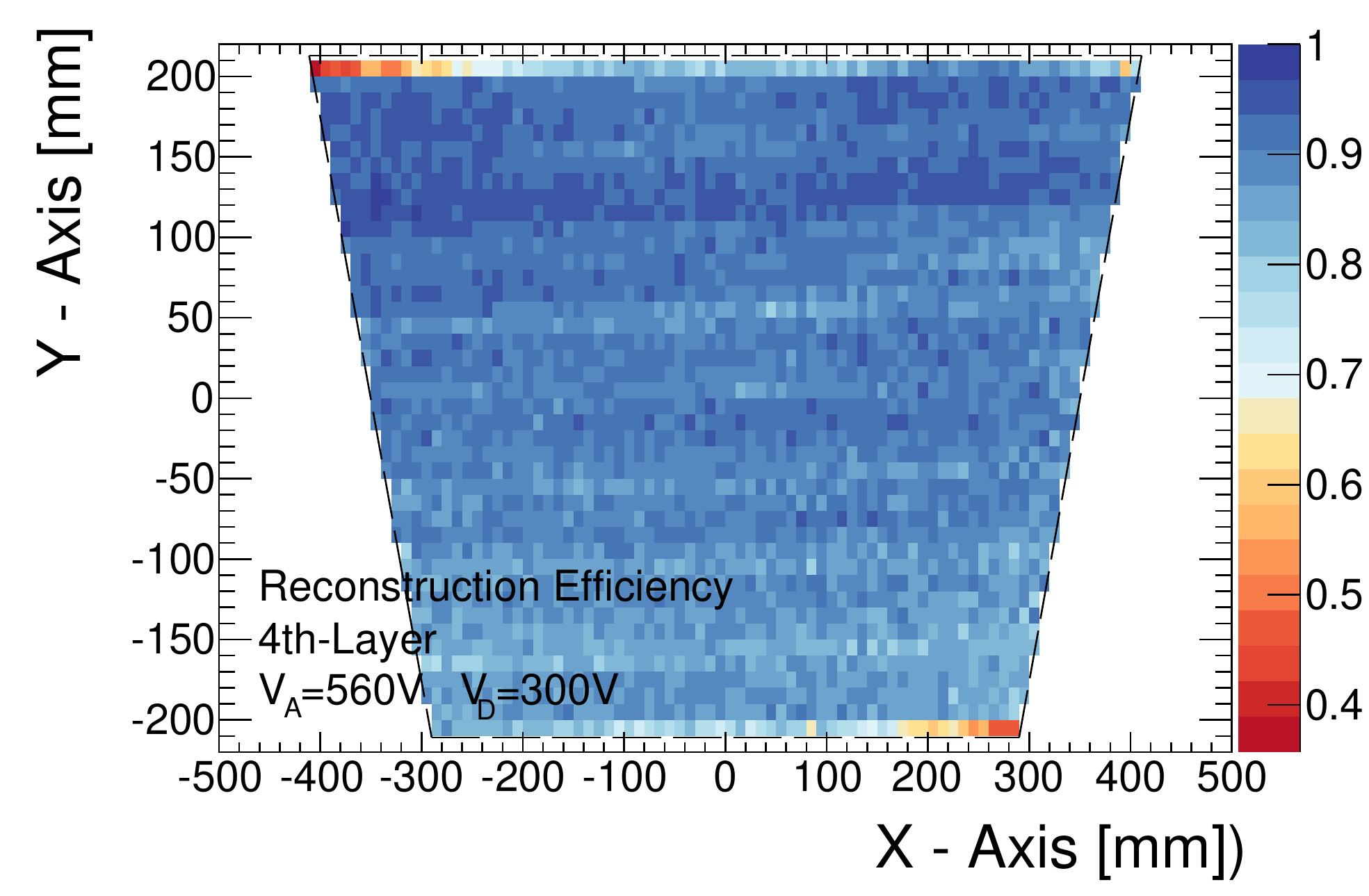}
        \includegraphics[width=0.49\textwidth]{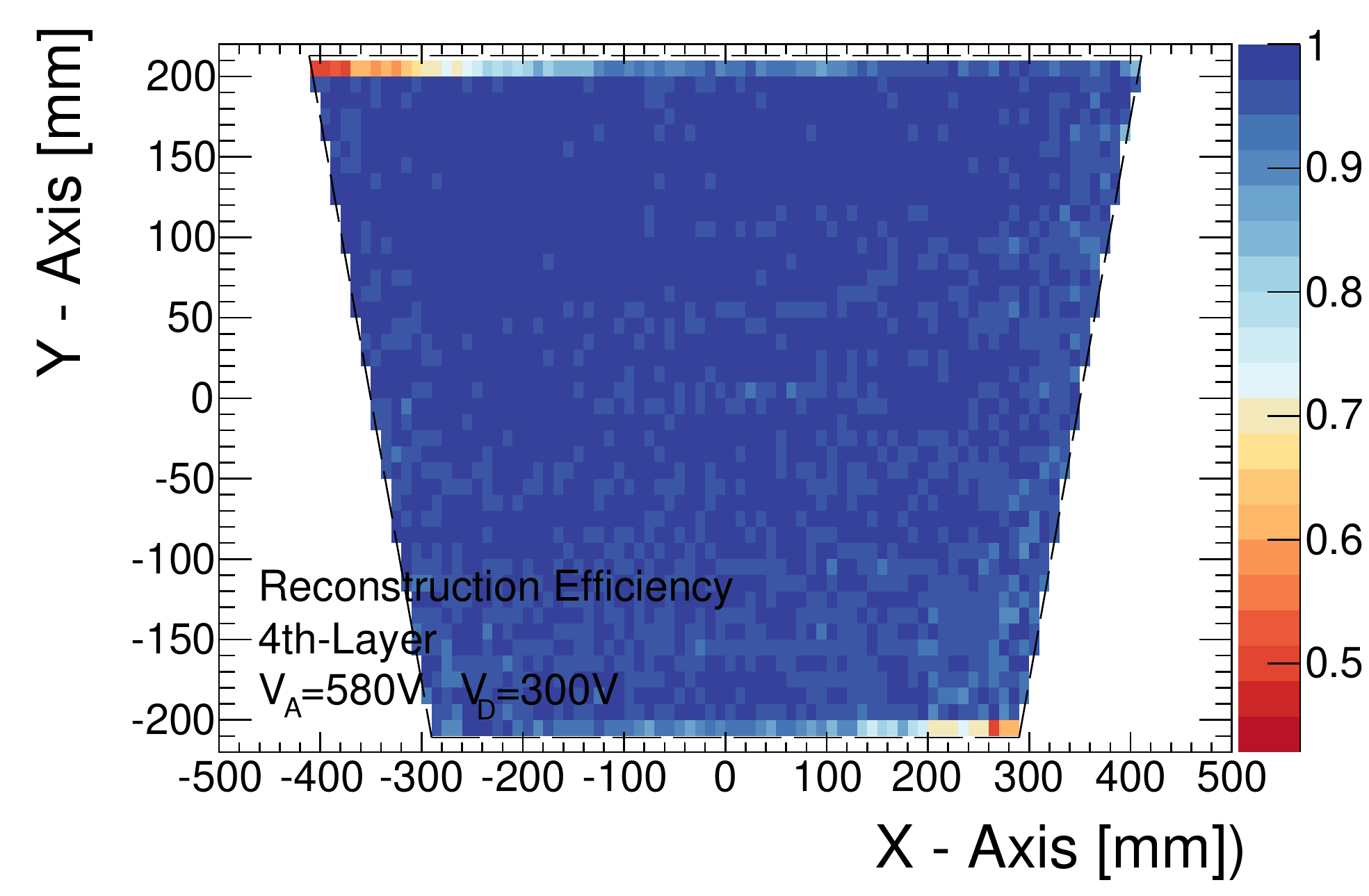}
        \caption{Two-dimensional cluster reconstruction efficiency over the full detector surface of the 4th layer for $V_\mathrm{A}=560\,$V (left) and $V_\mathrm{A}=580\,$V (right)}
        \label{Fig:Efficiency2D}
    \end{center}
\end{figure}

The detector efficiency in the fourth layer for $V_\mathrm{A}=560\,$V and $V_\mathrm{A}=580\,$V is shown in a two-dimensional representation of the detector surface in Fig. \ref{Fig:Efficiency2D}. The inefficiencies at the boundaries are the results of inclined strips (28 on the top, and 17 on the bottom), which are shorter at the edge and results to the corners which are not covered by the parallel strips. While we observe a higher efficiency in the top part compared to the lower part of the detector for $V_\mathrm{A}=560\,$V, a full homogeneous detector efficiency is found for $V_\mathrm{A} = 580\,$V. 

\subsection{Gain homogeneity}

\begin{figure}[h]
\begin{minipage}[hbt]{0.49\textwidth}
	\centering
	\includegraphics[width=0.98\textwidth]{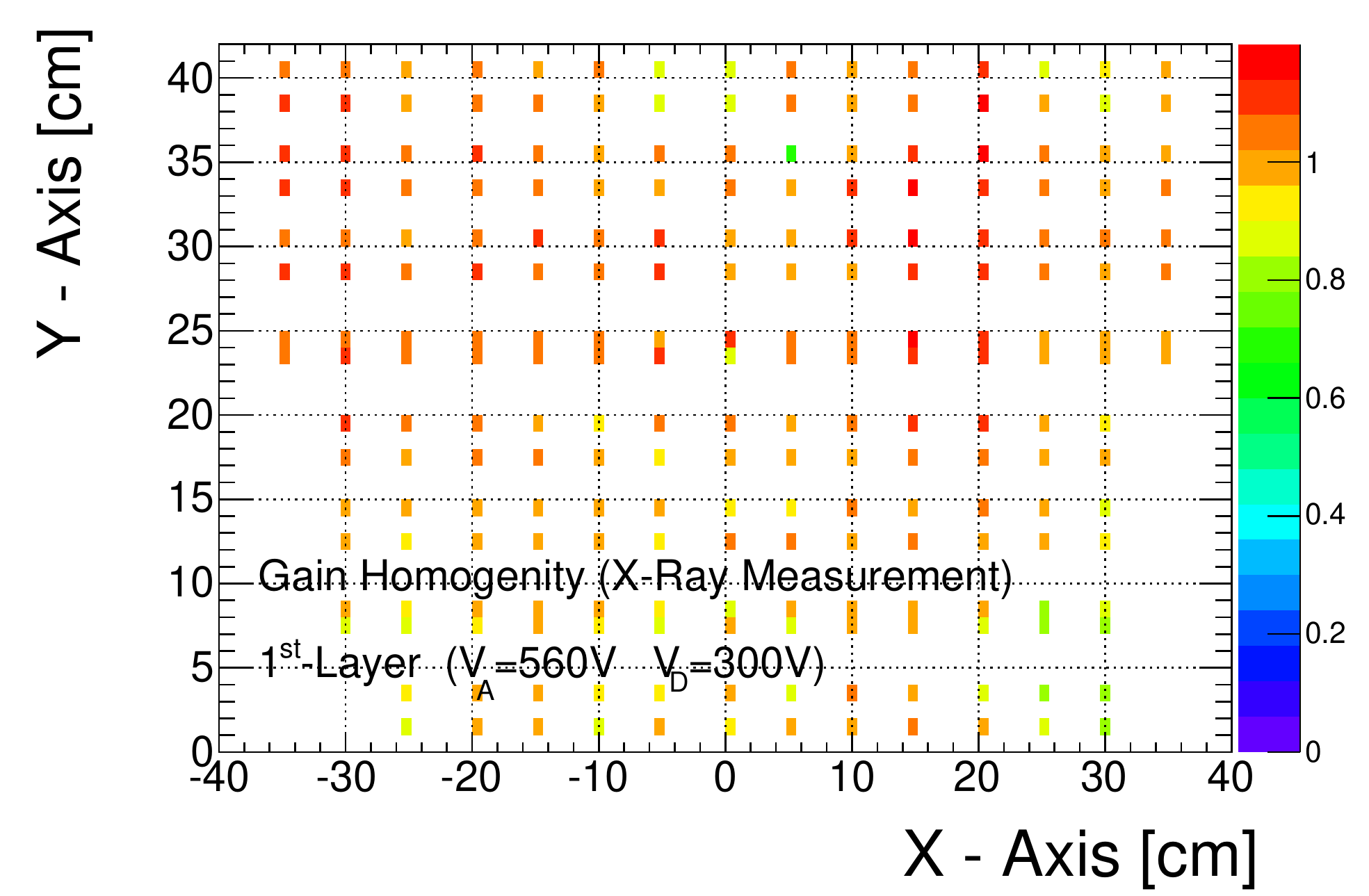}
	\caption{Normalized current map in the 1st MMSW layer as a result of the X-Ray irradiation.\vspace{0.4cm}}
	\label{Fig:Ran1}
\end{minipage}
\hspace{0.2cm}
\begin{minipage}[hbt]{.49\textwidth}
	\centering
	\includegraphics[width=0.98\textwidth]{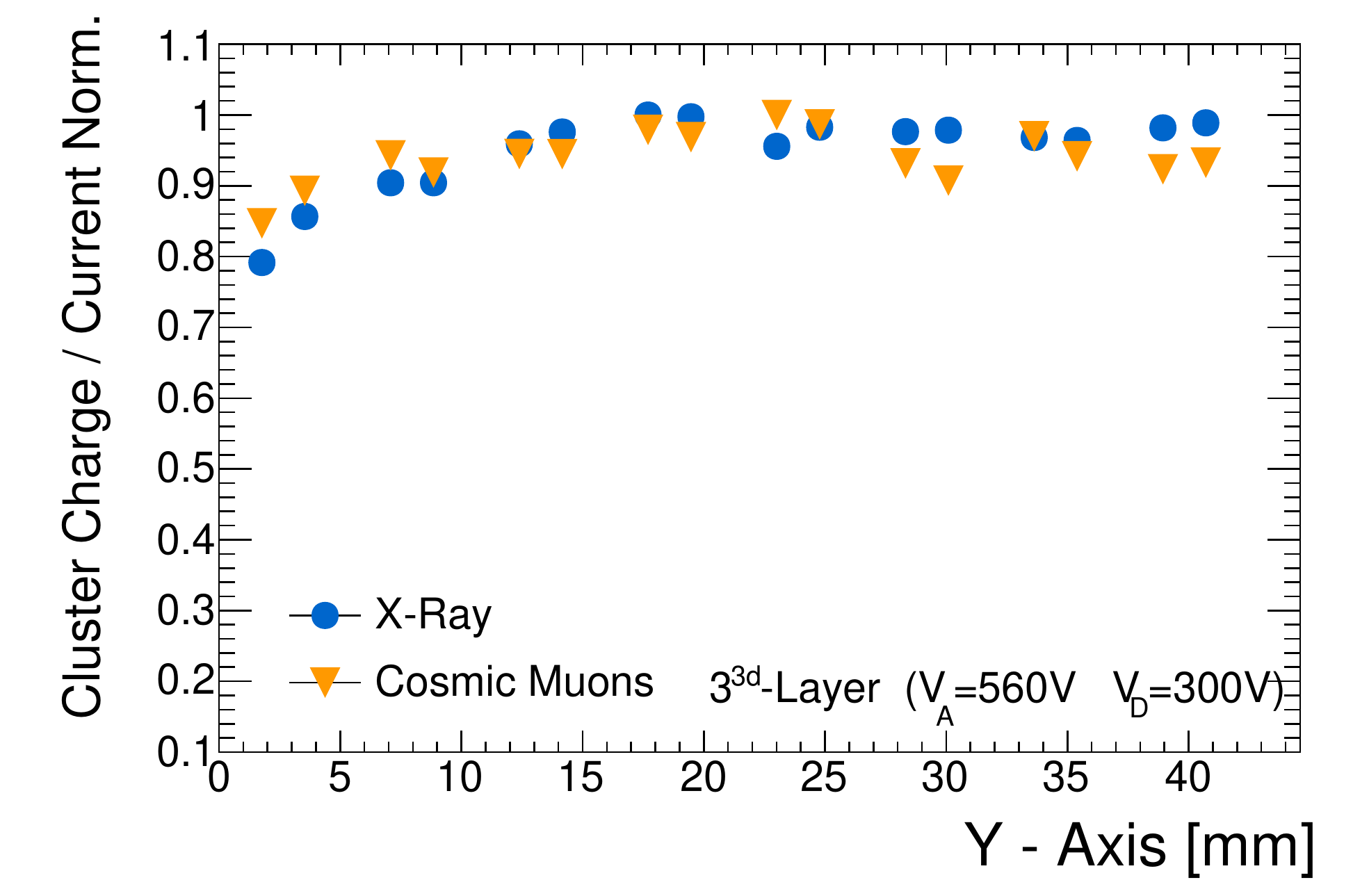}
	\caption{Normalized ratio of the average cluster charge, obtained with cosmic events, and the detector current, as a result of the X-ray irradiation, for the third readout layer.}
	\label{Fig:Ran2}
\end{minipage}
\end{figure}

The gain uniformity has also been tested by X-ray measurements. An Amptek\footnote{http://www.amptek.com/products/mini-x-ray-tube/} Mini-X Silver (Ag) source was used to irradiate the chamber in 228 different points over the detector surface. Setting the X-ray HV to 50\,kV and the X-ray current to 50$\,\mu$A, the detector current in each readout layer was monitored. Figure~\ref{Fig:Ran1} shows the normalized current in the first layer. The measured values are normalised to the average current. Points with very low current correspond to the edges of the active area of the chamber. A slightly higher amplification is observed in the upper part of the readout layer. The overall uniformity of the detector amplification is about 20\%, corresponding to a few microns uniformity of the amplification gap. This value is in agreement with the requirements of the ATLAS NSW project.

These results have been cross-checked against the cluster charge of cosmic events reconstructed in the same area that was previously irradiated by the X-ray source. Figure~\ref{Fig:Ran2} shows the comparison between X-ray and cosmic muon measurements. Both measurements are in good agreement, confirming the good uniformity of the amplification gap height.

\subsection{Spatial resolution in test beam measurements}

Test-beam measurements have been conducted in August 2014 at the MAMI accelerator facility at the Johannes Gutenberg-University Mainz for a study of the spatial resolution of the MMSW-1 detector. The MAMI accelerator delivers a quasi-continuous electron beam with energies up to $1.5\,$GeV.  For the measurements presented here the beam energy was set to $855$\,MeV. Following the previous studies, an Ar:$\mathrm{CO_2}$ gas mixture of 93:7 and drift- and amplification voltages of $V_\mathrm{D}=-300\,$V and $V_\mathrm{A}=550\,$V have been used. 

A small $10\times10$\,cm$^2$ micromegas chamber with two-dimensional readout and a spatial resolution of $70\,\mu\mbox{m}$ has been operated at a distance of $30\,$cm behind the MMSW detector. Details of this reference chamber are summarized in \cite{Lin:2014jxa}. This chamber served for reference measurements. 

\begin{figure}[t]
    \begin{center}
        \includegraphics[width=0.49\textwidth]{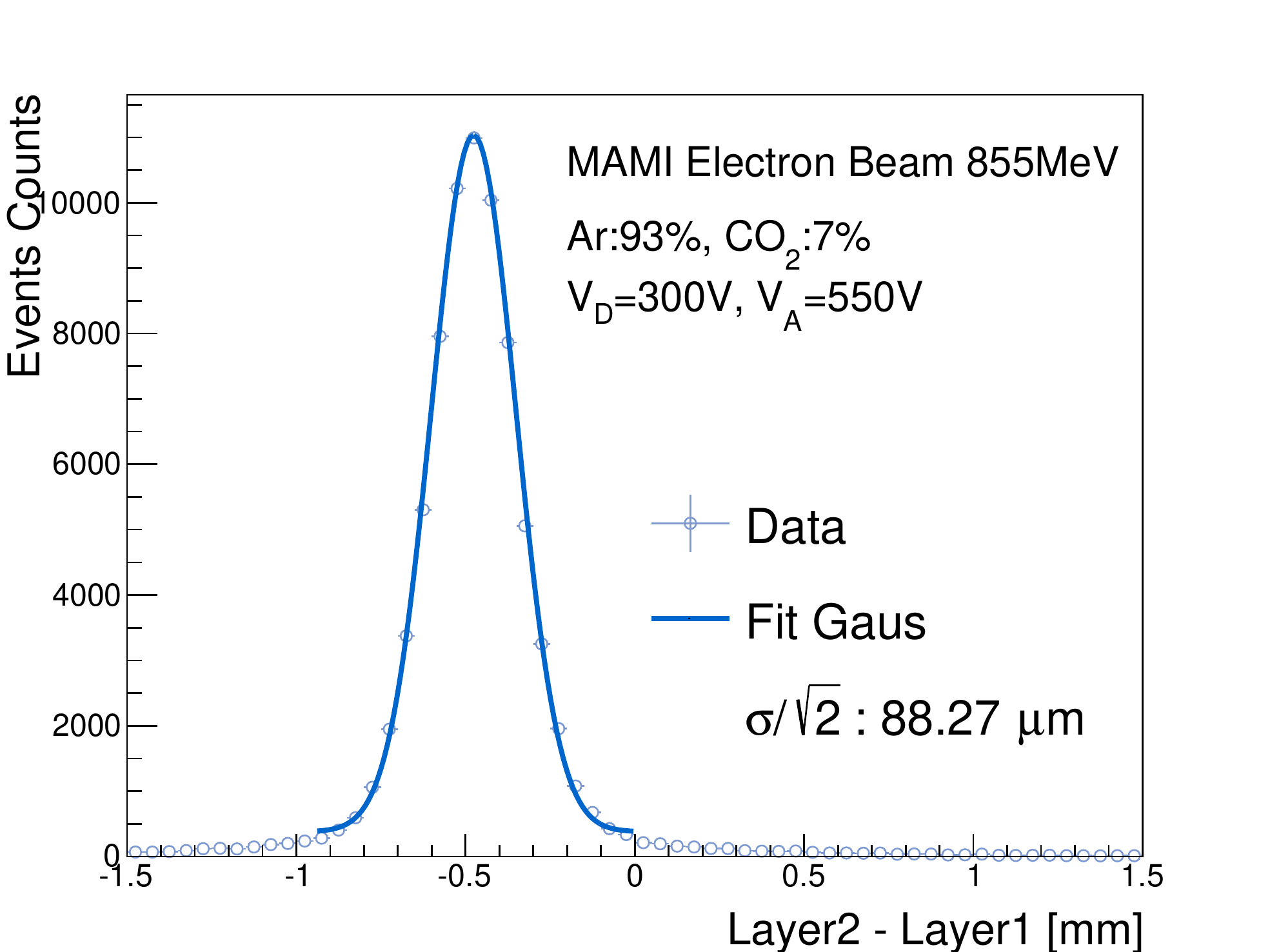}
        \includegraphics[width=0.49\textwidth]{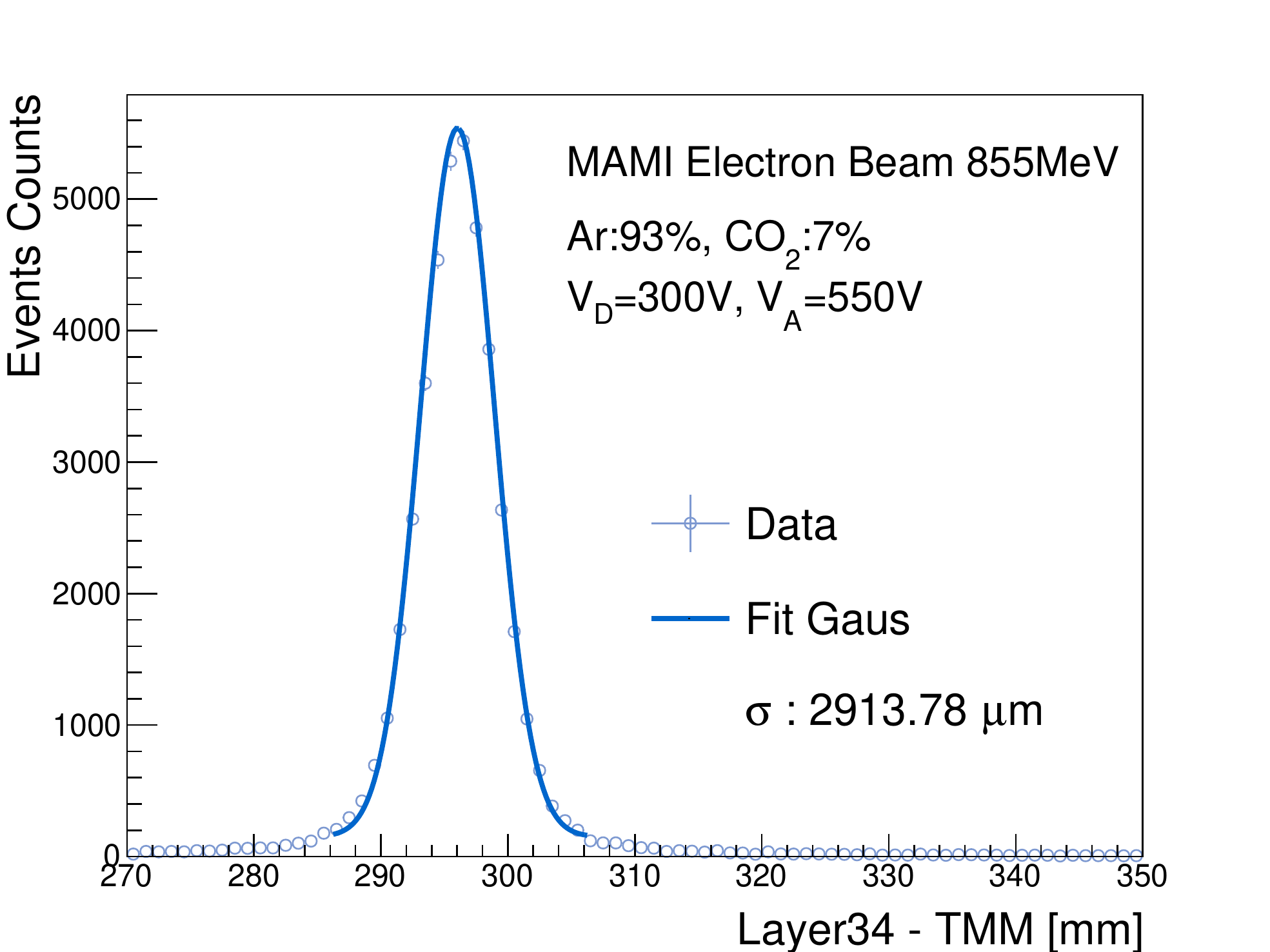}
        \caption{Difference of the reconstructed cluster position of the first and second layer (left) and difference of the y-coordinate position, reconstructed with the 3rd and 4th layer of the MMSW chamber and the reference chamber for test beam events (right).}
        \label{Fig:SpatialResolution}
    \end{center}
\end{figure}

We define the strip direction of the first and second layer of the MMSW as y axis. The spatial resolution in x direction of these two layers can therefore be estimated by comparing the reconstructed position of perpendicular incident particles in both layers on an event-by-event basis. Figure~\ref{Fig:SpatialResolution} (left) shows the difference of the reconstructed track positions in Layers 1 and 2. 
The width of the distribution $\sigma _\mathrm{Diff}$ is given by

\[ \sigma _\mathrm{Diff} = \sqrt{\sigma_{1}^2+ \sigma_{2}^2 + \sigma_\mathrm{BD}^2  + \sigma_\mathrm{MS}^2} \]

where $\sigma_{i}$ is the spatial resolution of the $i$-th layer, $\sigma_\mathrm{MS}$ is the contribution from multiple-scattering and $\sigma_\mathrm{BD}$ the contribution of the beam divergence. The offset of the observed distribution is an effect of the chamber orientation w.r.t. the beam. The contributions to the measured resolution from multiple scattering and the average opening angle of the electron beam are negligible compared to $\sigma_{i}$. The distribution can be described by a gaussian function\footnote{The gaussian accounts for more than 95\% of the events.}. Assuming that the intrinsic resolution is the same for both layers, we find a spatial resolution of $\sigma_i=$88~$\mu$m\footnote{Possible misalignment effects during the test beam measurements have not yet been accounted for.}, which is within the requirements of the ATLAS muon upgrade project.

The strips of the 3rd and 4th layer of the MMSW detector are inclined by $1.5^\circ$ and hence allow for the reconstruction of the y coordinate. The spatial resolution of the y-coordinate measurement via the 3rd- and 4th-layer was estimated via a simultaneous reconstruction of events in the small $10\times10$ cm reference chamber with two-dimensional readout. The difference in the reconstructed position of the small chamber and the y-position measurement of the MMSW is shown in Fig. \ref{Fig:SpatialResolution}. The width of this distribution has to be corrected for multiple scattering, which has been measured to be $1.8\,$mm, leading to a final y-reconstruction precision of the MMSW chamber of $\sigma _\mathrm{y} = 2.3\,$mm. Also this value is in accordance with the ATLAS Upgrade requirements.

\section{\label{sec:conclusion}Summary and Outlook}

The construction of two detectors based on the Micromegas technology in view of the ATLAS Muon Upgrade projects has been described. It should serve as a starting point for similar detector developments in the future. Each detector has an active area of $0.5\, \mathrm{m}^2$ and four planes of one-dimensional Micromegas readout structures. In two of the four planes the readout strips form a stereo angle of $\pm$1.5$^\circ$, allowing for the measurement of a second coordinate. Difficulties during the construction of the first detector, have been addressed and improved for the second version. 

First performance measurements based on cosmic rays, X-rays and test beam measurements at CERN and at the MAMI accelerator give results consistent with the ATLAS Muon System requirements. The response of the detector, operated with an Ar:CO$_2$ gas mixture, for different amplification voltages has been studied. An average signal reconstruction efficiency of $\ge$98\% per layer was found for an amplification voltage V$_\mathrm{A}$ $\ge$580~V, homogenous across the full detector. The intrinsic spatial resolution was determined in an electron beam of $855\,$MeV to be better than $100\,\mu \mathrm{m}$ in the precision coordinate and 2.3 mm in the second coordinate. 

The installation of one of the prototype detectors in the ATLAS experiment is currently being prepared. The main goal is to equip the detector with the final readout electronics that will be used for the ATLAS Upgrade project and to fully include the chamber in the ATLAS readout framework.

\section*{Acknowledgements}

We would like to acknowledge the close collaboration with the CERN PCB
workshop. In addition, we would like to thank A. Ochi (Kobe University) for providing the resistive kapton foils and the MAMI accelerator team, M. Hoeck and S. Caiazza for their help during our test-beam measurements. The contribution of the University of Mainz was supported by the Volkswagen Foundation and the German Research Foundation (DFG).

\bibliography{MMSWPrototype}{}
\bibliographystyle{atlasnote}

\end{document}